\documentclass{aa}  

\usepackage{graphicx}
\usepackage{placeins}
\usepackage{txfonts}
\newcommand{\micron}{$\mu$m}              

\usepackage[colorlinks=true,linkcolor=bluea,allcolors=blue]{hyperref}

\begin{document}

   \title{JWST MIRI reveals the diversity of nuclear mid-infrared spectra of nearby type-2 quasars}


   \author{C. Ramos Almeida\inst{1,2}, 
I. Garc\'ia-Bernete\inst{3},
M. Pereira-Santaella\inst{4},
G. Speranza\inst{4},
R. Maiolino\inst{5,6,7},
X. Ji\inst{5,6},
A. Audibert\inst{1,2}, 
P. H. Cezar\inst{1,2}, 
J. A. Acosta-Pulido\inst{1,2},
A. Alonso-Herrero\inst{3},
S. García-Burillo\inst{8},
O. González-Martín\inst{9},
D. Rigopoulou\inst{10,11},
C. N. Tadhunter\inst{12},
A. Labiano\inst{3,13},
N. A. Levenson\inst{14}
\and
F. R. Donnan\inst{10}
}

   \institute{Instituto de Astrof\' isica de Canarias, Calle V\' ia L\'actea, s/n, E-38205, La Laguna, Tenerife, Spain\\
              \email{cra@iac.es}
              \and Departamento de Astrof\' isica, Universidad de La Laguna, E-38206, La Laguna, Tenerife, Spain
\and Centro de Astrobiolog\' ia (CAB), CSIC-INTA, Camino Bajo del Castillo s/n, E-28692, Villanueva de la Cañada, Madrid, Spain
\and Instituto de Física Fundamental, CSIC, Calle Serrano 123, 28006 Madrid, Spain
\and Kavli Institute for Cosmology, University of Cambridge, Madingley Road, Cambridge, CB3 0HA, UK
\and Cavendish Laboratory - Astrophysics Group, University of Cambridge, 19 JJ Thomson Avenue, Cambridge, CB3 0HE, UK
\and Department of Physics and Astronomy, University College London, Gower Street, London WC1E 6BT, UK
\and Observatorio Astron\'omico Nacional (OAN-IGN)-Observatorio de Madrid, Alfonso XII, 3, 28014 Madrid, Spain
\and Instituto de Radioastronomía and Astrofísica (IRyA-UNAM), 3-72 (Xangari), 8701 Morelia, Mexico
\and Department of Physics, University of Oxford, Oxford OX1 3RH, UK
\and School of Sciences, European University Cyprus, Diogenes street, Engomi, 1516 Nicosia, Cyprus 
\and Department of Physics \& Astronomy, University of Sheffield, S3 7RH, Sheffield, UK
\and Telespazio UK for the European Space Agency, ESAC, Camino Bajo del Castillo s/n, 28692 Villanueva de la Cañada, Spain
\and Space Telescope Science Institute, 3700 San Martin Drive, Baltimore, MD 21218, USA
}

   \date{Received December 20, 2024; accepted March 30, 2025}

 
  \abstract
{Type-2 quasars (QSO2s) are active galactic nuclei (AGN) seen through a significant amount of dust and gas that obscures the central supermassive black hole and the broad line region. Here we present new mid-infrared spectra of the central kiloparsec of five optically-selected QSO2s at redshift z$\sim$0.1 obtained with the Medium Resolution Spectrometer (MRS) module of the Mid-Infrared Instrument (MIRI) aboard the James Webb Space Telescope (JWST). These QSO2s belong to the Quasar Feedback (QSOFEED) sample and they have bolometric luminosities of log L$_{\rm bol}$=45.5 to 46.0 erg~s$^{-1}$, global star formation rates (SFRs) that place them above the main sequence, and practically identical optical spectra in terms of spectral shape and [OIII] luminosity, but their nuclear mid-infrared spectra exhibit an unexpected diversity of both continua and features. They show: 1) 9.7 \micron~silicate features going from emission (strength of S$_{9.7}$=0.5) to relatively strong absorption (S$_{9.7}$=-1.0) and 18 and 23 \micron~silicates either in emission or flat (S$_{18}$=[0.2,0.0] and S$_{23}$=[0.1,0.0]). {\color {black} In addition, two of the QSO2s show absorption bands of CO, H$_2$O, and aliphatic grains, indicating different levels of nuclear obscuration across the sample.} 2) [NeV]/[NeII] ratios ranging from 0.1 to 2.1 and [NeIII]/[NeII] from 1.0 to 3.5, indicating different coronal line and ionizing continuum strengths. 3) Warm molecular gas masses of 1-4$\times$10$^7$ M$_{\sun}$ and warm-to-cold gas mass ratios of 1-2\%, with molecular gas excitation likely due to jet-induced shocks in the case of the Teacup (J1430+1339), and to UV heating and/or turbulence in J1509+0434. 4) Polycyclic aromatic hydrocarbon (PAH) emission features with equivalent widths ranging from <0.002 to 0.075 \micron, from which we measure a larger contribution from neutral molecules (PAH 11.3/6.2=1.3-3.4) and SFRs$\le$3-7 M$_{\sun}$~yr$^{-1}$. This unprecedented dataset allowed us to start exploring the role of various AGN and galaxy properties including ionizing continuum, obscuration, electron density, and jet-ISM interactions on some of the spectral differences listed above. Larger samples observed with JWST/MIRI are now required to fully understand the diversity of QSO2s' nuclear mid-infrared spectra.}

   
\titlerunning{The diversity of mid-infrared spectra of type-2 quasars}
\authorrunning{C. Ramos Almeida et al.}

   \maketitle
%

\section{Introduction}
\label{intro}

Type-2 quasars (QSO2s) are optically selected active galactic nuclei (AGN) with L$_{\rm [OIII]}>$10$^{8.3}L_{\sun}$ that show permitted emission lines with a full width at half maximum (FWHM) of $<$2000 km s$^{-1}$ \citep{Zakamska03,2008AJ....136.2373R}. They are the torus-obscured version of type-1 quasars, as revealed by polarimetry data of some QSO2s \citep{2005AJ....129.1212Z}, although in some cases obscuration might also come from galactic scales. Indeed, QSO2s have been proposed to be a dust-embedded phase during which AGN-driven outflows start clearing up gas and dust to eventually become unobscured quasars \citep{1988ApJ...325...74S,2009ApJ...696..891H}, making them ideal laboratories to study AGN feedback using different approaches and observations  \citep{Ramos22,Hervella23,Bessiere24,Girdhar24,Molyneux24,Speranza24,Ulivi24,Zanchettin25}. 

\begin{table*}
\small
\centering
  \caption{Properties of the QSO2s.} 
\begin{tabular}{lccccccccccc}
\hline
\hline
SDSS ID   & Short & SDSS & Scale & A$_{\rm V}$ & log L$_{\text{bol}}$ &  log L$_{1.4\text{GHz}}$ &  log M$_{\rm BH}$  & log\(\frac{L_{\rm bol}}{L_{\rm Edd}}\)  &  log M$_*$ & \multicolumn{2}{c}{SFR} \\
 & ID & z & (kpc/~\arcsec) & (mag) & (erg s$^{-1}$) & (W Hz$^{-1}$) & (M$_{\odot}$)  & & (M$_{\odot}$) & \multicolumn{2}{c}{(M$_{\odot}$yr$^{-1}$)}  \\
\hline
J101043.36+061201.4 & J1010 & 0.0977 & 1.807 & 1.1 & 45.6 & 24.4 & 8.4$\pm$0.8  &  -0.8$\pm$0.8 & 11.0$\pm$0.2 & 32 & 34 \\ 
J110012.39+084616.3 & J1100 & 0.1004 & 1.851 & 0.6 & 45.9 & 24.2 & 7.8$\pm$0.4  &   0.0$\pm$0.5 & 11.0$\pm$0.2 & 36 & 13 \\
J135646.10+102609.0 & J1356 & 0.1232 & 2.213 & 0.9 & 45.5 & 24.4 & 8.6$\pm$0.3  &  -1.0$\pm$0.4 & 11.3$\pm$0.2 & 73 & 1   \\
J143029.88+133912.0 & J1430 & 0.0851 & 1.597 & 0.9 & 45.8 & 23.7 & 8.2$\pm$0.4  &  -0.4$\pm$0.4 & 11.2$\pm$0.1 & 13 & 3 \\
J150904.22+043441.8 & J1509 & 0.1115 & 2.028 & 2.2 & 46.0 & 23.8 & 8.3$\pm$0.8  &  -0.2$\pm$0.8 & 10.9$\pm$0.3 & 36 & 3 \\
\hline
\end{tabular}
\tablefoot{A$_{\rm V}$ values were measured using H$_{\alpha}$/H$_{\beta}$ reported by \citet{Kong18} and the extinction law from \citet{Cardelli89}. The only exception is J1356, whose A$_{\rm V}$ value is unrealistically low (0.1 mag) and therefore we use the value measured from VLT/MUSE data instead, as in \citet{Zanchettin25}. Bolometric luminosities were calculated by applying the correction factor of 474 from \citet{Lamastra09} to the extinction-corrected [O~III]5007 \AA~luminosities from \citet{Kong18}, and 1.4 GHz luminosities and stellar masses are from \citet{Ramos22}. BH masses and Eddington ratios are from \citet{Kong18}. The last two columns are the global SFRs derived from total IR luminosities from \citet{Ramos22}, divided by 0.94 to convert them from Chabrier to Kroupa initial mass function (IMF), and the SFRs derived by \citet{Bessiere24} from spectral synthesis modelling of the SDSS spectra (3\arcsec$\sim$5-6 kpc) shown in Fig. \ref{figA1}.} 
\label{tab:sample} 
\end{table*}

Due to their high obscuration, the mid-infrared spectrum of QSO2s is a powerful tool to probe ionized and shocked regions lying behind dust clouds and dusty regions themselves. The 9.7 \micron~silicate feature strength (S$_{\rm 9.7}$) correlates with AGN type and X-ray gas column density (n$_{\rm H}$; \citealt{Shi06,Hatziminaoglou15}), with obscured, type-2 AGN generally showing it in shallow absorption and unobscured, type-1 AGN in weak emission or absent. However, examples of type-2 AGN with silicate features in emission and of type-1 with silicates in shallow absorption have been reported in the literature (see e.g. \citealt{Hatziminaoglou15,Mariela20}), and constituted one of the motivations for the development of clumpy torus models (see \citealt{Ramos17review} for review). For QSO2s, \citet{Zakamska08} reported silicate features in absorption or absent for a sample of 12 QSO2s with redshifts between 0.2 and 0.7 observed with the InfraRed Spectrograph (IRS) of the Spitzer Space Telescope. In a later work, \citet{Zakamska16} reported Spitzer/IRS S$_{\rm 9.7}$ measurements for a sample of 46 type-2 AGN at a median redshift of z=0.17, which range between 0.3 and -2.5 (median value of -0.30), with negative values corresponding to absorption features.

The mid-infrared range also contains emission lines from several ions of different species covering a large range of ionization potentials (IPs), making it possible to characterize the ionization state of the gas and the hardness of the radiation field \citep{Groves08,Zakamska08}. Some of these lines have IPs from $\sim$100 eV up to 200 eV (e.g. [FeVIII]$_{\rm 5.45}$, [MgVII]$_{\rm 5.50}$, [MgV]$_{\rm 5.61}$, [NeVI]$_{\rm 7.65}$, and [NeV]$_{\rm 14.3}$ \micron), known as coronal lines. They have been always associated with the presence of AGN because of the energetic radiation field required \citep{Ardila06,Ardila11,Ramos09,Muller11}, although  [NeV]$_{\rm 14.3}$ and [NeVI]$_{\rm 7.7}$ have been recently detected in James Webb Space Telescope (JWST) observations of the starburst galaxy M83 \citep{Hernandez25}. In general they are stronger in the nuclear region of active galaxies, but they have been detected up to distances of hundreds of parsecs in radio-quiet AGN \citep{Ramos17,Ardila20} and of a few kiloparsecs in radio galaxies \citep{Tadhunter87,Tadhunter88,Worrall12}. Shocks induced by jet-interstellar medium (ISM) interactions have been proposed as a plausible mechanism to explain their detection on these large scales in both radio-loud and radio-quiet AGN. 

Other spectral features of interest in the mid-infrared spectra of QSO2s are the H$_2$ rotational transitions, which probe molecular gas at hundreds of kelvin \citep{Rigopoulou02,Dasyra11,Guillard12,Togi16,Pereira22}, filling the gap between the hot molecular gas traced by the near-infrared ro-vibrational H$_2$ lines \citep{Ramos17,Ramos19,Speranza22} and the cold molecular gas that can be probed with e.g. the carbon monoxide (CO) lines detectable in the millimeter and sub-millimeter regime \citep{Jarvis20,Ramos22,Audibert23,Molyneux24}. In AGN and star forming galaxies, the emission of warm molecular gas is often enhanced by shocks and turbulence \citep{Ogle10,Pereira22,Kristensen23,GarciaBernete24b,Riffel25}, and thus, studying its mass, distribution, and kinematics is key to advance our understanding of e.g. jet and wind-ISM interactions. 

In addition, several polycyclic aromatic hydrocarbon (PAH) bands are also found in the mid-infrared, and they are commonly used as tracers of star formation \citep{1998ApJ...498..579G,1999AJ....118.2625R,2016ApJ...818...60S}. These PAH bands have been detected in the nuclear spectra of nearby Seyfert galaxies up to distances as close as tens of parsecs from the nucleus \citep{2013MNRAS.429.2634S,Alonso14,Alonso16,2014ApJ...780...86E,Esparza18,2022A&A...666L...5G,GarciaBernete24b}. Nevertheless, it is well-known that AGN radiation and shocks can modify the structure of the aromatic molecules and/or destroy the ionized and smaller grains, resulting in the neutral and larger molecules that produce the 11.3 and 17 $\mu$m features being enhanced relative to the short-wavelength PAHs (6.2, 7.7, and 8.6 $\mu$m; \citealt{Smith07,2010ApJ...724..140D,GarciaBernete22,GarciaBernete22a}). At quasar luminosities >10$^{\rm 45}$ erg~s$^{-1}$, it is not clear yet whether the PAH features are suitable or not as probes of recent star formation, since even the more resilient molecules might be destroyed in the nuclear regions \citep{2022ApJ...925..218X,Ramos23}.

Here we present James Webb Space Telescope (JWST) mid-infrared spectra of the first QSO2s at z<0.1 observed with the Medium Resolution Spectrometer (MRS; \citealt{Wells15}) of the Mid-Infrared Instrument (MIRI; \citealt{Glasse15,Rieke15,Wright23}). This dataset makes it possible to quantify the properties of nuclear gas and dust of QSO2s with unprecedented spatial and spectral resolution and sensitivity, and it has revealed an unexpected diversity of continuum shapes and spectral features. 
In the following we assume a cosmology with H$_0$=70 km~s$^{-1}$ Mpc$^{-1}$, $\Omega_m$=0.3, and $\Omega_{\Lambda}$=0.7.

\section{Sample, observations, and data reduction}
\label{observations}

The five QSO2s studied here are part of the Quasar Feedback (\href{http://research.iac.es/galeria/cra/qsofeed/}{QSOFEED}) sample \citep{Ramos22,Pierce23,Bessiere24}, selected from the \citet{2008AJ....136.2373R} compilation of narrow-line AGN to have L$_{\text{[OIII]}} > 10^{8.5}$L$_{\odot}$ and redshift of z<0.14. 
{These five QSO2s} have redshifts {0.09 $\leq$ z $\leq$ 0.12},  bolometric luminosities of log L$_\text{bol} = 45.5-46.0$ erg~s$^{-1}$, radio luminosities of log L$_{1.4\text{GHz}}$=23.7-24.4 W~Hz$^{-1}$, optical extinctions of A$_{\rm V}$=0.6-2.2 mag, and stellar masses of log M$_*$ = 10.9-{11.3} M$_\odot$ (see Table \ref{tab:sample}). They have black hole (BH) masses of {log M$_{\rm BH}\sim$7.8-8.6} M$_{\sun}$ and Eddington ratios ranging from log L$_\text{bol}/L_\text{Edd}$={-1 (J1010 and J1356)} to 0 (J1100), with J1430 and J1509 having intermediate values of -0.4 and -0.2 (see Table \ref{tab:sample}).

  \begin{figure*}
   \centering
  \includegraphics[width=1.75\columnwidth]{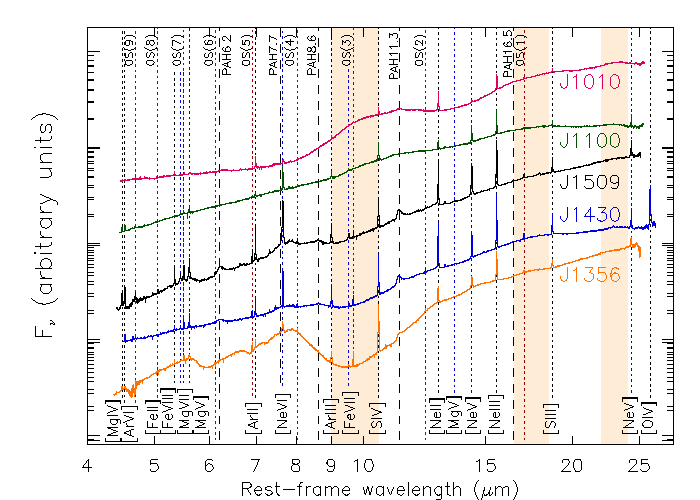}
   \caption{JWST/MIRI nuclear spectra of the central 0.7-1.3 kpc of the QSO2s, scaled in the Y-axis using a multiplicative factor to sort them out according to S$_{\rm 9.7}$, and smoothed using a boxcar of 5. For reference, the flux densities at 20 \micron~measured from these nuclear spectra are 464, 249, 144, 114, and 75 mJy for J1100, J1010, J1430, J1356, and J1509. The amorphous silicate features at 9.7 and 18 \micron~and the crystalline silicate feature at 23 \micron~are highlighted with the light orange areas. The most intense emission lines and PAHs are labelled, with the high and low-ionization atomic lines shown with blue and black dotted lines, molecular lines with red dotted lines, and PAH bands with black dashed lines.}
              \label{fig1}%
    \end{figure*}

These five targets are representative of the most gas-rich QSO2s in the QSOFEED sample, as they have total molecular gas masses of M$_{\rm H_2}$=4-18$\times$10$^{9}$ M$_{\sun}$ and evidence for cold molecular outflows detected from Atacama Large Millimeter Array (ALMA) CO(2-1) observations \citep{Ramos22,Audibert23}. As shown in Figure 1 in \citet{Speranza24}, these QSO2s have the highest AGN luminosities and the most extreme ionized gas kinematics within the QSOFEED sample. The most remarkable differences between these QSO2s are found when looking at the optical and radio morphologies: J1010 and J1430 are early-type galaxies in the pre- and post-coalescence stages of a galaxy interaction, {J1356 is an on-going merger,} and J1100 and J1509 are seemingly undisturbed spiral galaxies with bars \citep{Ramos22,Pierce23}. The global SFRs derived from total IR luminosities (see Table \ref{tab:sample}) place the QSO2s above the star-formation main sequence and, in the case of the two spirals, a significant proportion of star formation seems to be in the outer part of spiral arms, as the SFRs derived from spectral synthesis modelling of the optical SDSS spectra (central 3\arcsec$\sim$5-6 kpc) of the QSO2s are significantly smaller, especially in the cases of J1356 and J1509 (see Table \ref{tab:sample}). 
At centimeter and/or sub-millimeter wavelengths, the 0.2-0.25\arcsec~resolution Very Large Array (VLA) and ALMA data of J1430 and J1509 show extended continuum emission (hundreds of parsecs), whilst J1010, J1100, and J1356 appear compact \citep{Jarvis19,Ramos22}. On larger spatial scales, J1356 and J1430 show radio emission extending up to several kiloparsecs \citep{Jarvis19,Speranza24}. The five QSO2s are ``radio-quiet'' but they all show a radio excess that cannot be accounted for by star formation \citep{Jarvis19,Ramos22}. See \citet{Harrison24} and \citet{Njeri25} for further discussion on ``radio-AGN'' definitions.

The QSO2s were observed from May {2024 to January 2025} with the integral field unit of JWST/MIRI, the MRS, as part of Cycle 2 General Observer (GO) Program 3655 (PI: C. Ramos Almeida). The total exposure times range from 1.36 to 5.49 hours using FASTR1 reading mode for the brightest target (J1100) and SLOWR1 for the others. We used single pointing and a 4-point dither pattern for all the targets and obtained background observations of half the exposure time of the science observations, using a 2-point dither pattern. In order to avoid having residual persistence in the background observations, they were obtained before their corresponding science target. We refer the reader to the Program Information webpage of Program \href{https://www.stsci.edu/jwst/science-execution/program-information?id=3655}{{GO 3655}} for further details on the MRS observations.

The MIRI/MRS data cover the spectral range 4.9-28.1 $\rm{\mu m}$, with a spectral resolution of R$\sim$3700-1300 \citep{2021A&A...656A..57L,Argyriou23}. The data are split in four channels that cover the following ranges: [4.9-7.65] $\mu$m (Channel 1; hereafter Ch1), [7.51-11.71] $\mu$m (Ch2), [11.55-18.02] $\mu$m (Ch3), and [17.71-28.1] $\mu$m (Ch4). Combining these four channels and the three grating settings available (short, medium, and long), there are 12 different wavelength bands. The angular resolutions measured for these 12 bands range from $\sim$0.3\arcsec~to 0.8\arcsec~and the field-of-view (FOV) of the different channels goes from 3.2\arcsec$\times$3.7\arcsec~in Ch1 to 6.6\arcsec$\times$7.7\arcsec~in Ch4 (see Table 1 in \citealt{Esparza25}). The data were reduced using the JWST Science Calibration Pipeline (v 1.13.4), with the context 1216 for the calibration References Data System \citep{Labiano16}, adding a few extra steps to better remove hot and dead pixels by applying a mask and interpolating the continuum. The background subtraction was done channel by channel for each science cube using background average spectra generated with the dedicated background observations of each sub-band. We also applied a residual fringe correction to each science spectrum using the script \textit{rfc1d-utils}. See \citet{Pereira22} and \citet{GarciaBernete24} for more details on the data reduction.

\section{Results}
\label{results}

\begin{table*}
\small
\centering
\caption{Measurements of key spectral features detected in the nuclear spectra of the QSO2s.}
\begin{tabular}{lccccccc}
\hline
\hline
QSO2 & \multicolumn{2}{c}{S$_{9.7}$} & S$_{18}$ & S$_{23}$ & [NeIII]$_{15.6}$/[NeII]$_{12.8}$ & [NeV]$_{14.3}$/[NeII]$_{12.8}$  &  H$_2$S(5)/S(1)  \\
     & Obs. & No PAH  & Obs.& Obs. &     &   &           \\
\hline
J1010 &  0.49$\pm$0.01      & 0.49        & 0.22$\pm$0.02       & 0.10$\pm$0.02 & 1.02$\pm$0.09 & 0.11$\pm$0.15    & 0.40$\pm$0.19 \\
J1100 &  0.11$\pm$0.01      & 0.11        & 0.11$\pm$0.02       & 0.07$\pm$0.04 & 3.48$\pm$0.06 & 2.07$\pm$0.06  &  {1.10$\pm$0.42} \\
J1356 & -1.04$\pm$0.02      &-0.95        & 0.06$\pm$0.03       & \dots         & 3.22$\pm$0.07 & 1.22$\pm$0.08  &  1.25$\pm$0.06 \\
J1430 &{-0.29$\pm$0.09} & {-0.26} & {0.18$\pm$0.06} & 0.07$\pm$0.02 & 1.79$\pm$0.06 & 0.55$\pm$0.07  &  {0.43$\pm$0.06} \\
J1509 &  -0.19$\pm$0.10     & -0.05       & 0.06$\pm$0.04       & \dots         & 1.67$\pm$0.03 & 0.99$\pm$0.04  &  0.97$\pm$0.06 \\
\hline
\end{tabular}
\tablefoot{Columns 2, 3, 4, and 5 list the 9.7 $\mu$m silicate feature strength measured from the spectra before and after subtracting the PAH component, and of the 18 and 23 $\mu$m silicate features. Errors of the PAH-subtracted S$_{9.7}$ are the same as those of the observed S$_{9.7}$, but do not include the uncertainty associated with PAH subtraction. Columns from 6 to 9 correspond to emission line ratios aimed to characterize the strength and hardness of the ionizing continuum, of the coronal line emission relative to low-IP lines, and molecular gas excitation.} 
\label{tab1} 
\end{table*}

We extracted the nuclear spectra of the QSO2s from all the sub-channels assuming that they are point sources (e.g. $\sim$0.4\arcsec~at 11\,$\mu$m). 
We fitted a 2D Gaussian to the nuclear emission of each spectral channel to determine the size of the aperture (from $\sim$0.3\arcsec~to 0.8\arcsec~with increasing wavelength), extracted the flux in that aperture, and applied an aperture correction, 
as described in Appendix A of \citet{GarciaBernete24}. Therefore, considering the range of aperture sizes, at the redshift of the targets (z$\sim$0.1), the nuclear spectra correspond to a physical scale of $\sim$0.7--1.3 kpc. The spectra of the 12 bands were stitched together to make them coincide in the overlapping regions. Small scaling factors were applied to some of the sub-channels to match their overlapping region with the adjacent, shorter wavelength sub-channel (e.g. we multiplied Ch2m by a factor to match Ch2s). These factors range from f=0.94 to 1.06 and they were calculated as the ratio of the average fluxes of the overlapping region of two contiguous channels (in the previous example, f=<Ch2s>/<Ch2m>).

The diversity of spectral shapes and features shown by the nuclear mid-infrared spectra of these QSO2s is apparent, as it can be seen from Fig. \ref{fig1}. This is at odds with their optical spectra, which are very similar in terms of spectral shape (see Fig. \ref{figA1} in Appendix \ref{appendix}) and [OIII] luminosity.  Nevertheless, some key optical emission line ratios (e.g. HeII$\lambda$4686/H$_{\beta}$), already indicated differences among them (see Section \ref{discussion:excitation}). In the following subsections we present and analyze various features measured from the nuclear mid-infrared spectra of the QSO2s.

\subsection{Silicate features}
\label{silicates}

  \begin{figure}
   \centering
  \includegraphics[width=0.93\columnwidth]{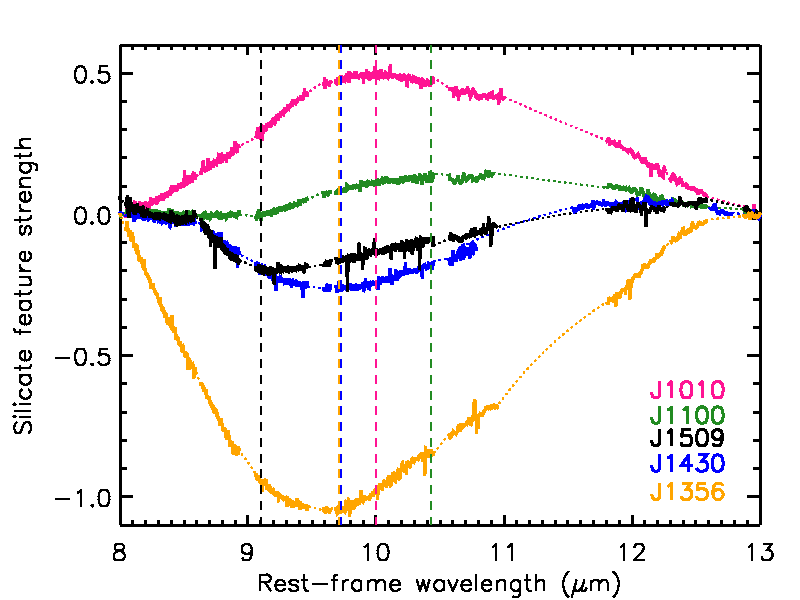}
   \caption{Silicate feature strength of the QSO2s. The peak wavelengths are indicated with vertical dashed lines of matching colors, which correspond to 10.0 and 10.4 \micron~for J1010 and J1100 (emission features) and {9.7, 9.7, and 9.1 for J1356, J1430,} and J1509 (absorptions). Dotted lines correspond to the masked spectral regions for removing the contribution of PAH features and narrow emission lines.}
              \label{fig2}%
    \end{figure}

We quantify the 9.7 \micron~silicate features detected in the spectra shown in Fig. \ref{fig1} with their strength, which we define as S$_{9.7}$ = {\color {black} ln(F$_{\rm 9.7}$)$-$ln(F$_{\rm cont}$). F$_{\rm 9.7}$ is the peak flux of the silicate feature and F$_{\rm cont}$ is the corresponding continuum flux at the peak wavelength, calculated from a spline fit of the continuum.}
We then repeated the same procedure after removing the broad PAH features, which in the case of J1430 and J1509 are strong, using an infrared fitting tool that models the PAH features on top of a continuum generated using a differential extinction model \citep{Donnan24}. The values of S$_{9.7}$ before and after PAH removal are reported in Table \ref{tab1}. A negative (positive) value indicates a feature in absorption (emission). In Fig. \ref{fig2} we show the silicate strength measured for the observed features (i.e., without removing the PAH emission), and the peak wavelengths, which range from 9.1 $\mu$m in the case of J1509 to 10.4 $\mu$m in J1100. These variations in the peak wavelength are due to different dust composition and/or size \citep{Mariela20,GarciaBernete22cons,Omaira23,Ulises24}. 

\begin{figure*}
   \centering
  {\par\includegraphics[width=0.9\columnwidth]{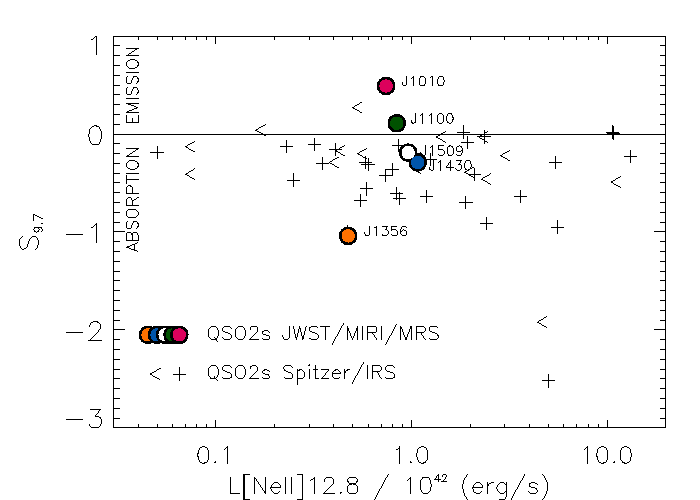}
  \includegraphics[width=0.9\columnwidth]{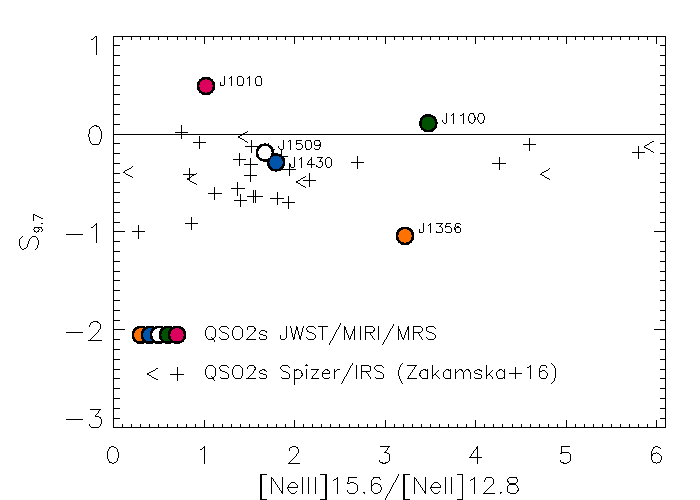}}
   \caption{Observed S$_{9.7}$ versus [NeII] luminosity (left) and [NeIII]/[NeII] (right). The values and upper limits measured from Spitzer/IRS spectra of type-2 AGN \citep{Zakamska16} are shown as crosses and arrows, and the values measured for the five QSO2s with MIRI/MRS observations as circles of different colors. The horizontal line at S$_{9.7}$=0 indicates the division between emission (positive) and absorption (negative) features.}
              \label{fig3}
    \end{figure*}

In J1010 we observe the strongest silicate emission feature, for which we measure S$_{9.7}$=0.49{$\pm$0.01}, followed by J1100, with S$_{9.7}$=0.11$\pm$0.01. These measurements do not change after removing the PAH features (see Table \ref{tab1}). Silicates in emission are uncommon in QSO2s \citep{Zakamska16}, although measurements from Spitzer/IRS spectra correspond to larger spatial scales {than those studied here}, and hence foreground extinction from the host galaxy might be contributing to observe silicates in absorption {more often} \citep{Lacy07,Levenson07,Gonzalez13}. In the case of J1430 and J1509 we measure S$_{9.7}$=-0.29$\pm$0.09 and -0.19$\pm$0.10 before PAH subtraction (see Fig. \ref{fig2}), and -0.26 and -0.05 afterwards, consistent with shallow absorption, more typical of local QSO2s (see Fig. \ref{fig3}; \citealt{Zakamska16}). Indeed, J1509 was included in the Spitzer/IRS sample studied by \citet{Zakamska16}, who reported an observed value of S$_{9.7}$=-0.26, slightly deeper than the nuclear value measured here from the MIRI spectrum. Finally, the other extreme of the sample is the merging QSO2 J1356, which shows the deepest silicate feature: S$_{9.7}$=-1.04{$\pm$0.02 and -0.95} before and after PAH subtraction. This value is among the deepest silicate features reported by \citet{Zakamska16}, as shown in Fig. \ref{fig3}.

We also quantified the strength of the 18 and 23 $\mu$m silicate features (S$_{18}$ and S$_{23}$) following the same procedure as for S$_{9.7}$ (see Table \ref{tab1}). In the case of the former we measure peak wavelengths at $\sim$17-18 $\mu$m, and all the features are in emission, with strengths ranging from S$_{\rm 18}$=0.22 (J1010) to 0.06 (J1356 and J1509). The 23 \micron~band of crystalline silicates \citep{Spoon06,Spoon22} is detected in weak emission in the spectra of J1010, J1100, and J1430, {with peak wavelengths at $\sim$23 $\mu$m and strengths of S$_{\rm 23}$=0.07-0.1} (see Fig. \ref{fig1} and Table \ref{tab1}). The strength of this band 
is correlated with the amorphous silicate strength (e.g. the 9.7 and 18 \micron~features; \citealt{Spoon22}), with J1010 showing the most prominent 23 \micron~band, followed by J1100 and J1430. In the case of the 18 and 23 \micron~bands, the change from an emission to an absorption feature happens at higher obscuration than in the case of the 9.7 \micron~feature because their wavelengths are longer. This might be the reason behind the flat 23 \micron~bands of J1356 and J1509, and their weak emission 18 \micron~bands (see below). The 18 and 23 $\mu$m bands have been detected in absorption in ultra-luminous infrared galaxies (ULIRGs; \citealt{Spoon06,Spoon22,Donnan23}).

\begin{figure}
   \centering
  \includegraphics[width=0.95\columnwidth]{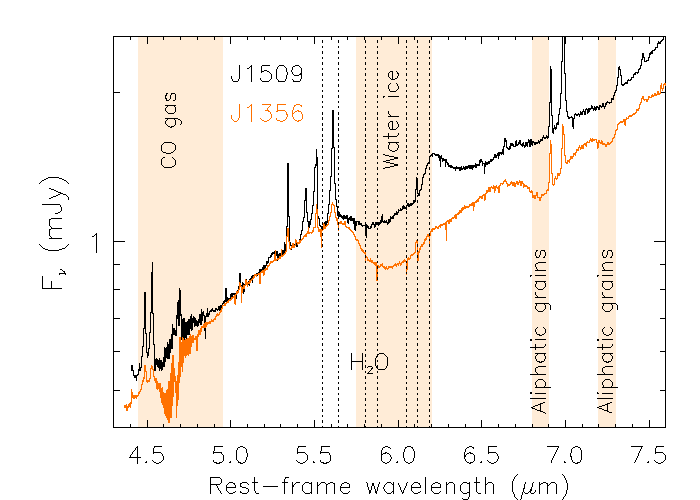}
   \caption{Nuclear spectra of J1356 and J1509, {\color {black} normalized at 5.4 \micron, showing several absorption features: the ro-vibrational 4.67 \micron~CO band, H$_2$O lines between $\sim$5.5 and 6.2 \micron, the 6 \micron~water ice band, and the 6.85 and 7.25 \micron~aliphatic grain bands. Some of the most prominent H$_2$O lines are indicated with dotted vertical lines.}}
              \label{fig:ices}
    \end{figure}

The silicate feature strengths that we measure from the nuclear spectra of the QSO2s are indicative of moderate nuclear obscuration, explaining the lack of {\color {black} absorption features including CO, H$_2$O, and aliphatic grains} in J1010, J1100, and J1430. These {\color {black} bands} are mainly detected in ULIRGs \citep{Spoon01,Spoon22} and obscured Seyfert galaxies \citep{GarciaBernete24,Gonzalez25}. On the contrary, in the case of J1356 and J1509, which have the highest nuclear column densities measured from CO together with J1430 (log N$_{\rm H}^{\rm CO}\sim$23 cm$^{-2}$; see Section \ref{discussion_silicates}), we detect {\color {black} the ro-vibrational $\sim$4.45-4.95 \micron~CO band (hereafter 4.67 \micron~CO band), several H$_2$O lines from $\sim$5.5 to 6.2 \micron~\citep{GarciaBernete24soda}, the $\sim$5.8-6.2 \micron~water ice (6 \micron~water ice), and the 6.85 and 7.25 \micron~hydrogenated amorphous carbon grains (hereafter aliphatic grains; see Fig. \ref{fig:ices} and \citealt{GarciaBernete24}). For the water ice} we measured peak wavelengths at 6.03 and 5.95 \micron~and observed strengths of S$_6$=-0.38$\pm$0.01 and -0.15$\pm$0.05 {\color {black} for J1356 and J1509}, respectively. This band is a mix of water, CO, CO$_2$, and other molecules, and its strength appears to be correlated with X-ray derived column densities \citep{GarciaBernete24}. All the {\color {black} absorption bands} detected in J1356 are deeper than those in J1509, indicating higher nuclear obscuration in the former.

\begin{table*}
\small
\centering
\caption{Properties of the [NeII]$_{\rm 12.8}$, [NeV]$_{\rm 14.3}$, and H$_2$0-0S(5) emission lines derived from the fits with Gaussian components.}
\begin{tabular}{lccccccccc}
\hline
\hline
QSO2 &\multicolumn{3}{c}{[NeII]$_{\rm 12.8}$} &\multicolumn{3}{c}{[NeV]$_{\rm 14.3}$} &  \multicolumn{3}{c}{H$_2$0-0S(5)} \\
      & L$_{\rm [NeII]}$ &  FWHM & v$_s$ & L$_{\rm [NeV]}$ &  FWHM & v$_s$  & L$_{\rm S(5)}$ &  FWHM & v$_s$       \\
      & (10$^{\rm 41}$ erg~s$^{\rm -1}$) & (km~s$^{-1}$) &  (km~s$^{-1}$)& (10$^{\rm 41}$ erg~s$^{\rm -1}$) & (km~s$^{-1}$) &  (km~s$^{-1}$) & (10$^{\rm 41}$ erg~s$^{\rm -1}$) & (km~s$^{-1}$) & (km~s$^{-1}$) \\
\hline
J1010  & {3.50$\pm$0.22} &  410$\pm$10  &  0$\pm$3    &  {0.14$\pm$0.03}   & 390         & 40$\pm$20   & {0.18$\pm$0.01} & 340$\pm$30 & -30$\pm$10  \\
       & {3.92$\pm$0.19} & 1220$\pm$60  &  10$\pm$10  &  {0.66$\pm$0.05}   & 1600$\pm$100& 170$\pm$50  & \dots               & \dots      & \dots       \\
J1100  & {3.78$\pm$0.16} &  360$\pm$10  &  0$\pm$3    &  {8.88$\pm$0.15}   & 339$\pm$4   & 28$\pm$1    & {0.53$\pm$0.06} & 510$\pm$60 & -10$\pm$20   \\
       & {4.63$\pm$0.17} & 1360$\pm$60  & 60$\pm$20   &  {8.57$\pm$0.19}   & 1470$\pm$40 & 120$\pm$10  & \dots               & \dots      & \dots       \\
J1356  & {1.30$\pm$0.05} &  300$\pm$10  & 200$\pm$5   &  {0.81$\pm$0.08}   & 360$\pm$25  & 140$\pm$6   & {0.90$\pm$0.02} & 470$\pm$10 & -10$\pm$4   \\
       & {1.95$\pm$0.06} &  290$\pm$10  & -200$\pm$3  &  {0.73$\pm$0.04}   & 220$\pm$10  & -270$\pm$4  & \dots               & \dots      &  \dots      \\
       & {1.50$\pm$0.08} & 1640$\pm$90  & -100$\pm$30 &  {4.23$\pm$0.09}   & 1450$\pm$30 & 14$\pm$8    & \dots               & \dots      &  \dots      \\
J1430  & {4.65$\pm$0.17} &   320$\pm$6  & 0$\pm$1     &  {3.04$\pm$0.11}   & 420$\pm$8   & 56$\pm$2    & {0.50$\pm$0.01} & 500$\pm$10 & 20$\pm$4    \\
       & {6.12$\pm$0.15} &  900$\pm$20  & -13$\pm$4   &  {2.94$\pm$0.10}   & 1150$\pm$40 & 80$\pm$7    & \dots               & \dots      &  \dots      \\
J1509  & {5.57$\pm$0.09} &  386$\pm$4   & 0$\pm$1     &  {3.07$\pm$0.08}   & 420$\pm$8   & 7$\pm$2     & {0.59$\pm$0.02} & 390$\pm$10 & 25$\pm$5    \\
       & {4.05$\pm$0.09} & 1390$\pm$40  & -140$\pm$10 &  {6.48$\pm$0.09}   & 1560$\pm$30 & -220$\pm$8  & \dots               & \dots      & \dots       \\
\hline
\end{tabular}
\tablefoot{The peak of the narrow component of [NeII] has been set as zero velocity (v$_s$=0 km~s$^{-1}$), so v$_s$ of any other component is relative to it. In J1356, where two narrow components were fitted to the atomic lines, the average of their two peak wavelengths corresponds to zero velocity. FWHMs are corrected from instrumental broadening, calculated following Eq. 1 in \citet{Argyriou23}. FWHMs without errors have been fixed.}
\label{tab:kin} 
\end{table*}

\subsection{High and low-ionization emission lines}
\label{lines}

Before fitting the emission lines, we corrected the spectra shown in Fig. \ref{fig1} from extinction using the A$_{\rm v}$ values reported in Table \ref{tab:sample}
and the local ISM extinction curve of \citet{Chiar06}. These extinction values have modest impact on the emission lines, except for the case of the rotational line H$_2$0-0S(3) because of its proximity to the silicate feature \citep{Pereira14,Donnan24,Davies24}. 
We modelled the mid-infrared emission lines with {up to three} Gaussians and a local continuum, {defined as a one-degree polynomial fitted to emission-line free regions of varying size blue and red-wards of the emission line.} Additional kinematic components might be required in some cases, but for the sake of simplicity here we limit them to a maximum of {three} (see Table \ref{tab:kin}). A detailed study of the gas kinematics exploiting the integral field capabilities of MIRI/MRS will be the subject of a forthcoming paper.

We used two Gaussian components to reproduce the majority of the line profiles of the low- and high-ionization atomic lines, which generally correspond to a narrow component of FWHM$\sim$300--400 km~s$^{-1}$ associated with the narrow-line region (NLR) and to a broader, blue- or red-shifted component of FWHM$\sim$900--1600 km~s$^{-1}$, representative of turbulent gas, likely associated with non-circular motions (see Table \ref{tab:kin}). In the case of J1356, three Gaussian components were necessary to reproduce the atomic lines: two narrow ones of FWHM$\sim$200-360 km~s$^{-1}$ and one broad of FWHM$\sim$1500--1600 km~s$^{-1}$. Similar fits were done for the recombination lines detected in the near-infrared spectrum of this QSO2 \citep{Zanchettin25}.

From the analysis of optical and near-infrared integral field spectroscopic data of these QSO2s we know that they all have ionized gas outflows detected in [OIII]5007 \AA~\citep{Harrison14,Venturi23,Bessiere24,Speranza24,Ulivi24}, and in the case of J1356, J1430, and J1509, also in Pa$\alpha$ and [SiVI]1.963\footnote{Tentative [SiVI] outflow detection in the case of J1356.} \micron~\citep{Ramos17,Ramos19,Zanchettin25}. The most extreme kinematics detected with JWST/MIRI correspond to J1509, which shows a prominent component of FWHM$\sim$1400-1600 km~s$^{-1}$, blueshifted by 140-220 km~s$^{-1}$ in both the low- and high-ionization atomic emission lines. This is in good agreement with the nuclear kinematics of the coronal line of [SiVI]1.963 (IP=167 eV) reported by \citet{Ramos19} using data from the Gran Telescopio Canarias (GTC; FWHM=1460 km~s$^{-1}$, v$_s$=-100 km~s$^{-1}$). {J1356 also shows a broad component of FWHM$\sim$1500--1600 km~s$^{-1}$, blueshifted by 100 km~s$^{-1}$ in the case of [NeII] and consistent with zero velocity in the case of [NeV]. \citet{Zanchettin25} reported redshifted Pa$_{\alpha}$ and Br$_{\gamma}$ components of FWHM$\sim$1300-1400 km~s$^{-1}$ and tentative outflow detection in [SiVI] based on the large FWHM of the only gaussian fitted, of FWHM$\sim$900 km~s$^{-1}$. Finally, for J1430} \citet{Ramos17} reported FWHM=1600 km~s$^{-1}$ and v$_s$=-77 km~s$^{-1}$ for the broad component of [SiVI]1.963 using data from the Very Large Telescope (VLT), which is significantly broader and more blueshifted than the broad components reported here for both the high- and low-ionization atomic lines (see Table \ref{tab:kin}). 

For all the QSO2s {but J1356} we find that the FWHM of the broad component of [NeV] is larger than that of the [NeII] line, which could be indicating that the coronal lines probe outflowing gas closer to the AGN. Indeed, there is a well-known correlation between the ionization potential of the emission lines and their FWHM, usually interpreted as a stratification of the line-emitting regions \citep{Ardila04,Ardila11,Muller11}.

One of the most striking differences between the spectra of the five QSO2s is the diversity of emission line ratios. The [NeIII]$_{\rm 15.6}$/[NeII]$_{\rm 12.8}$ ratio, which is used to characterize the gas ionization state and radiation field hardness \citep{Groves08}, is much higher in J1100 and J1356 (3.5$\pm$0.1 and 3.2$\pm$0.1) than in the other three QSO2s (1-1.8; see Table \ref{tab1} and right panel of Fig. \ref{fig3}). For comparison, in the right panel of Fig. \ref{fig3} we also show the [NeIII]/[NeII] ratios reported by \citet{Zakamska16} for their sample of nearby type-2 AGN, including QSO2s, observed with Spitzer/IRS. The majority of their values range from 0 to 2, with a handful of targets showing higher values of between 3 and 6, all of them showing silicate features in shallow absorption (S$_{\rm 9.7}$>-0.5; see Fig. \ref{fig3}). For J1509 \citet{Zakamska16} reported a ratio of [NeIII]/[NeII]=1.39, slightly lower than the value we measure from the MIRI nuclear spectrum. Using data of hundreds of AGN and star-forming galaxies with Spitzer/IRS spectra, \citet{Pereira10} reported a median value of [NeIII]/[NeII]=2 for QSOs, of 0.8-1.2 for Seyfert galaxies, of 0.3 for low-ionization emission regions (LINERs), and of 0.2 for star-forming galaxies.

\begin{figure}
\centering
\includegraphics[width=0.9\columnwidth]{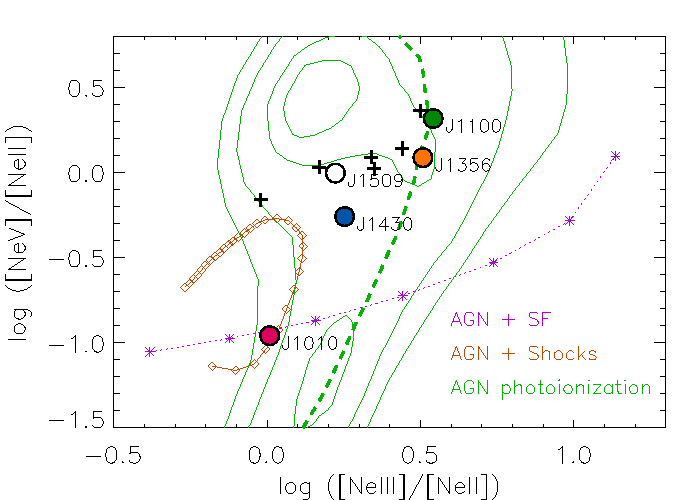}
\caption{[NeV]/[NeII] versus [NeIII]/[NeII] diagram. The five QSO2s are shown with the same symbols as in Fig. \ref{fig3} and the six GATOS Seyfert galaxies from \citet{Zhang24a} as black crosses. The green dashed curve is the AGN photoionization model from \citet{Feltre16} with gas metallicity Z=0.017, density of gas cloud n$_H$=10$^3$ cm$^{-3}$, dust-to-metal mass ratio $\xi_d$=0.3, UV spectral index $\alpha$=-1.7, and log U varying from -1.5 to -4.5 from top to bottom. Green contours are the full grid of AGN models from \citet{Feltre16} with Z>1/3 Z$_{\odot}$. Purple asterisks are AGN+SF models with 90\% of star formation contribution to H$_{\beta}$, with log U increasing from -3 to 0 from left to right. Brown diamonds are AGN+shocks models with 90\% of shock contribution to H$_{\beta}$, with shock velocity increasing from 200 to 1000 km~s$^{-1}$ counterclockwise, pre-shock density of 100 cm$^{-3}$, and transverse magnetic field strength = 1 $\mu$G. Adapted from Fig. 5 in \citet{Feltre23}. See also \citet{Hermosa24}.} 
\label{models}
\end{figure}

The nuclear spectrum of J1100 also shows strong emission from high-ionization lines (e.g. [NeV]$_{\rm 14.3}$, [NeVI]$_{\rm 7.63}$, [MgV]$_{\rm 5.61}$, [MgVII]$_{\rm 5.50}$, [FeVIII]$_{\rm 5.45}$, {\color {black} and [MgIV]$_{\rm 4.49}$,} which have IPs ranging from {\color {black} 80} to 186 eV), relative to low-ionization lines such as [NeII]$_{\rm 12.8}$ and [SIV]$_{\rm 10.51}$ (IPs of 21 and 35 eV). J1356, J1430, and J1509 also show strong coronal lines, but comparable to or weaker than their low-ionization emission lines, and J1010 has extremely faint coronal emission considering its quasar luminosity. This is quantified in Table \ref{tab1} and Fig. \ref{models} by means of the line ratio [NeV]$_{\rm 14.3}$/[NeII]$_{\rm 12.8}$. J1100 shows the highest value, of {2.1$\pm$0.1}, and J1010 the lowest, of {0.1$\pm$0.2}, coinciding with the limit of [NeV]/[NeII]$\ge$0.1 used by \citet{Inami13} to classify galaxies as AGN-dominated in the mid-infrared. J1356, J1430, and J1509 have intermediate values of this ratio, {of 0.6-1.2} (see Fig. \ref{models}). For comparison, the median ratios reported by \citet{Zakamska08} and \citet{Pereira10} for different QSO samples observed with Spitzer/IRS spectra are $\sim$1-1.5, while for different samples of Seyfert galaxies observed with either Spitzer/IRS or the Infrared Space Observatory (ISO) it ranges between 0.5 and 1 \citep{Sturm02,Tommasin08,Pereira10}. More recently, using JWST/MIRI MRS observations of a sample of six type-2 Seyferts from the Galactic Activity, Torus, and Outflow Survey (GATOS), \citet{Zhang24a} reported median nuclear ratios of 1.2 and 1.5 for [NeV]/[NeII] and [NeVI]/[NeII] respectively. One of the galaxies in their sample, NGC\,3081, has strong coronal emission, showing similar ratios as J1100. The [NeVI]/[NeII] ratios of the QSO2s measured from the nuclear spectra range from 0.1$\pm$0.1 (J1010) to 1.9$\pm$0.1 (J1100). 

\begin{table*}
\small
\centering
\caption{Density values derived from the [NeV]$_{\rm 14.3/24.3}$ ratio, the optical trans-auroral lines, and the [SII]$\lambda\lambda$6716,6731 \AA~doublet.}
\begin{tabular}{lccccc}
\hline
\hline
QSO2 & [NeV]$_{\rm 14.3/24.3}$ & \multicolumn{2}{c}{log n$_e^{\rm [NeV]}$} & log n$_e^{\rm [TR]}$ & log n$_e^{\rm [SII]}$  \\
   & & T$_e$=10$^4$ K    & T$_e$=2$\times$10$^4$ K &             &         \\
   & & (cm$^{-3}$) & (cm$^{-3}$)     & (cm$^{-3}$) &   (cm$^{-3}$)             \\
\hline
J1010 & >2.89$^*$     &  >4.07                  & >4.33                   & 4.58$\pm$0.03           & 2.95$\pm^{0.04}_{0.05}$ \\
J1100 & 2.32$\pm$0.05 & 3.88$\pm$0.02           & 4.14$\pm$0.02           & 3.99$\pm^{0.07}_{0.08}$ & 2.92$\pm$0.05 \\
J1356 & 1.00$\pm$0.30 & 2.8$\pm^{0.5}_{\dots}$  & 3.1$\pm^{0.5}_{\dots}$ & 3.21$\pm$0.15           & 2.49$\pm^{0.09}_{0.12}$ \\     
J1430 & {1.40$\pm$0.34} & {3.4$\pm^{0.2}_{0.5}$}    & {3.7$\pm^{0.2}_{0.5}$} & 3.24$\pm^{0.05}_{0.30}$ & 2.97$\pm^{0.05}_{0.06}$ \\
J1509 & 1.73$\pm$0.13 & 3.64$\pm^{0.05}_{0.09}$ & 3.90$\pm^{0.05}_{0.09}$ & 3.41$\pm^{0.11}_{0.21}$ & 2.80$\pm^{0.18}_{0.27}$ \\
\hline
\end{tabular}
\tablefoot{[NeV]$_{\rm 14.3/24.3}$ measured from the nuclear mid-infrared spectra and corresponding n$_e$ values derived with {\it PyNeb} (v 1.1.19; \citealt{Luridiana15}) for T$_e$=10$^4$ and 2$\times$10$^4$ K. {The values measured for J1356 do not have lower limit on the errors because [NeV]$_{\rm 14.3/24.3}$=0.70 does not permit to constrain n$_e$ at those temperatures (see Fig. \ref{density}).} The last two columns are the n$_e$ values derived from the optical trans-auroral ratios (involving [SII]$\lambda\lambda$4068,4076 \AA, [SII]$\lambda\lambda$6716,6731 \AA, [OII]$\lambda\lambda$3726,3729 \AA, and [OII]$\lambda\lambda$7319,7331 \AA) reported by \citet{Speranza24} and \citet{Bessiere24}, and from the [SII]$\lambda\lambda$6716,6731 \AA~doublet. * Value of the [NeV] ratio derived from the upper limit at 2$\sigma$ measured for [NeV]$_{\rm 24.3}$.}
\label{tab:density} 
\end{table*}

The optical spectra of J1100, shown in Fig. \ref{figA1}, shows [FeVII]$\lambda$6087 \AA~emission (IP=99 eV), which is also present in J1356 and J1509, marginal in J1430, and absent in J1010. The JWST nuclear spectra then reveal coronal emission from lines with IPs$\lesssim$200 eV in all the QSO2s including J1010 (albeit very weak in this target), and more importantly, a huge range of high-to-low ionization ratios (see Table \ref{tab1}). 

\subsection{Electron density}
\label{electron_density}

  \begin{figure}
   \centering
  \includegraphics[width=0.9\columnwidth]{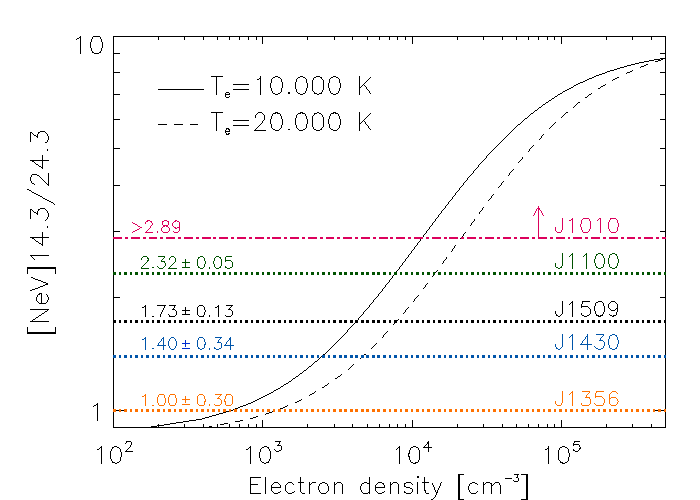}
   \caption{The density-sensitive ratio of [NeV]$_{\rm 14.3/24.3}$ as a function of electron density modelled with {\it PyNeb} (version 1.1.19; \citealt{Luridiana15}). Solid and dashed lines correspond to electron temperatures of 10$^4$ and 2$\times$10$^4$ K, respectively. The values of the flux ratio measured from the nuclear spectra are shown as dotted lines of different colors, and the upper limit measured for J1010 as dot-dashed line.}
              \label{density}%
    \end{figure}

The [NeV]$_{\rm 14.3}$ and [NeV]$_{\rm 24.3}$ emission lines can be used to measure the gas electron density (n$_e$), since their ratio is largely insensitive to temperature and extinction variations \citep{Alexander99}. Using data from Spitzer/IRS of hundreds of nearby galaxies, \citet{Pereira10} reported values of the [NeV]$_{14.3/24.3}$ ratio of 1.0-1.1 for QSOs and Seyferts. These are similar to the ratios reported for Seyfert galaxies using data from ISO by \citet{Sturm02}, of 1.1-1.3, and to the one we measure here for J1356, of 1.0$\pm$0.3 (see Table \ref{tab:density}). 
For an electron temperature (T$_e$) of 10$^4$ K, this ratio corresponds to log n$_e\sim$2.8 cm$^{-3}$ (see Fig. \ref{density}). For the other QSO2s we measure higher nuclear [NeV]$_{14.3/24.3}$ ratios, shown in Table \ref{tab:density} and Fig. \ref{density}. The highest {value} corresponds to J1010, for which we do not detect [NeV]$_{\rm 24.3}$ (see Fig. \ref{fig1}) and we use an upper limit at 2$\sigma$ to estimate its [NeV]$_{\rm 14.3/24.3}$ ratio of >2.89, which results in log n$_e$>4.07 cm$^{-3}$. {For J1100, J1430, and J1509 the [NeV]$_{14.3/24.3}$ ratios range from 1.4 to 2.3 and the corresponding densities, from log  n$_e$=3.4 to 3.9 cm$^{-3}$ (see Table \ref{tab:density} for the individual values including errors, which are larger in the case of J1356 and J1430). }
In Table \ref{tab:density} we also report the values of n$_e$ when T$_e$={2$\times$10$^4$ K} is used instead, as the electron temperatures measured for the QSO2s using the ratio [OIII]$\lambda$4363/[OIII]$\lambda$5007 \AA~from optical SDSS spectra range between $\sim$10,000 and 20,000 K (see also \citealt{Speranza22,Speranza24,Bessiere24}). 

{The electron densities derived from the [NeV]$_{14.3/24.3}$ ratio} are much higher than expected for coronal gas emitting in the optical (log n$_e\sim$1.2-1.7 cm$^{-3}$) according to the radiation-pressure-dominated model of \citet{Dopita02}, and they are in good agreement with the values estimated from the optical trans-auroral lines (two pairs of [SII] and [OII] doublets in the optical; see Table \ref{tab:density}). In the case of J1010, the electron density estimated from the trans-auroral lines, of log n$_e$=4.58{$\pm$0.03 cm$^{-3}$}, is higher than the [NeV]$_{\rm 24.3}$ critical density, of log n$_c$=4.42 cm$^{-3}$. We discuss the implications of these results in Section \ref{discussion:density}. Finally, Table \ref{tab:density} also shows the electron densities calculated using {\it PyNeb} (version 1.1.19; \citealt{Luridiana15}) and the ratios [SII]$\lambda$6731/$\lambda$6716 \AA~and [OIII]$\lambda$4363/[OIII]$\lambda$5007 \AA~measured from the SDSS spectra shown in Fig. \ref{figA1}, but with stellar continuum fitted by \citet{Bessiere24} subtracted. The corresponding electron densities are significantly lower than those derived from [NeV]$_{14.3/24.3}$ and the trans-auroral lines.

\subsection{Warm molecular gas}
\label{warm}

    \begin{table*}
    \small
    \centering
    \caption{Parameters derived from the rotational diagrams of the QSO2s and from ALMA CO(2-1) observations.}
    \label{tab:h2}    
        \begin{tabular}{lccccccccccc}
            \hline 
            \hline
            QSO2 & Fit & $\chi^{2}$ & $\beta$ & $\rm{T_w}$ &$\rm{\log (<N_{H}^w>) }$ &  $\rm{M_w}$ & $\rm{T_h}$ & $\rm{\log (<N_{H}^h>)} $ & $\rm{M_h}$ & $\rm{M_{H_2}^{nuc} (CO)}$ & Warm-to-cold \\
            & & &  & (K) & ($\rm{cm^{-2}}$) & ($\rm{10^6 \, M_{\odot}}$) & (K) & ($\rm{cm^{-2}}$) &  ($\rm{10^6 \, M_{\odot}}$) & ($\rm{10^9 \, M_{\odot}}$) & ratio\\ 
            \hline
            J1010 & PL & 7.33 &  5.24  & \dots & 20.60 &  {10.5}  & \dots &   \dots    & \dots       & 0.78  & {0.013} \\
                  & 2T & 0.12 & \dots  & 241   & 20.85 &  {18.6}  & 1017  &   {18.38}  & {0.06}  & \dots & {0.024} \\
            J1100 & PL & 2.48 &  4.45  & \dots & 20.64 &  {12.0}  & \dots &  \dots     & \dots       & 0.86  & {0.014} \\
                  & 2T & 3.54 & \dots  & 288   & 20.71 &  {14.2}  & 1010  &   {18.85}  & {0.20}  & \dots & {0.017} \\
            J1356 & PL & 7.01 &  4.33  & \dots & 20.66 &  {18.0}  & \dots &  \dots     & \dots       & 2.05  & {0.009} \\
                  & 2T & 15.7 & \dots  & 294   & 20.58 &  {15.0}  &  904  &  {19.09}   & {0.49}  & \dots & {0.008} \\
            J1430 & PL & 6.03 & 5.20   & \dots & 21.27 &  {38.2}  & \dots & \dots      & \dots       & 2.21  & {0.017} \\
                  & 2T & 10.5 & \dots  & 258   & 21.30 &  {41.0}  &{742}  & {19.57}    & {0.76}  & \dots & {0.019} \\
            J1509 & PL & 13.0 &  4.53  & \dots & 20.58 &  {12.6}  & \dots & \dots      & \dots       & 2.40  & {0.005} \\
                  & 2T & 6.50 & \dots  & 256   & 20.78 &  {20.0}  & 1107  & {18.68}    & {0.16}  & \dots & {0.008} \\   
                  \hline
            \end{tabular}
    \tablefoot{Results from the fit with the single power law temperature distribution (PL), which represents both the hot (h) and warm (w) {molecular} gas and assumes that the column density follows a power law (dN$_{H}\sim$T$^{\rm \beta}$dT), are shown in the same columns as those corresponding to the warm component of the two-temperature (2T) fits. The column densities correspond to average values within the PSF radius and they should be considered as lower limits. The last two columns correspond to the mass of molecular gas measured in the central kiloparsec of the QSO2s (r=0.5 kpc for J1100, J1356, J1430, and J1509 and r=0.7 kpc for J1010), calculated from the CO(2-1) ALMA observations presented in \citet{Ramos22} and assuming R$_{21}$=1 and $\alpha_{CO}$=4.35 M$_{\sun}$(K~km~s$^{-1}$~pc$^2$)$^{-1}$, and the corresponding warm-to-cold gas mass ratio.}
    \end{table*}

One of the key motivations for observing AGN of different properties with JWST/MIRI is the access that it provides to the rotational H$_2$ lines, which probe molecular gas at temperatures of a few hundred K, with unprecedented resolution and sensitivity \citep{Davies24,Esparza25,Riffel25}. As it can be seen from Fig. \ref{fig1}, the H$_2$ lines from 0-0S(9) to S(1) are clearly detected in {J1356 and J1509, which are the QSO2s} with the most intense H$_2$ emission at T>500 K (i.e. lines from S(4) to S(9); see Fig. \ref{fig1} and Tables \ref{tab:kin} {and \ref{tab:lines1}}). These lines are also detected in J1010 and J1430, but in J1100 the lines from S(6) to S(9) appear weak, possibly because of the strong continuum of this QSO2. The S(7) line at 5.511 $\mu$m is blended with [MgVII]$_{\rm 5.503}$, making it challenging to separate them. In J1010, J1356, and J1509 the two lines can be identified, with the S(7) being stronger than [MgVII]$_{\rm 5.503}$, whilst in J1100 and J1430 [MgVII]$_{\rm 5.503}$ dominates the blend. Based on the simplistic approach that we used here to model the emission line profiles, we find that we can reproduce the H$_2$ profiles of the five QSO2s with just one Gaussian. The FWHM of the H$_2$ lines ranges {between} $\sim$300 and 600 km~s$^{-1}$ (see Tables \ref{tab:kin} {and \ref{tab:lines1})}, consistent with the FWHM of the narrow components fitted to the atomic lines. The broadest profiles are those of J1430, of 500-600 km~s$^{-1}$ {for} all the H$_2$ lines from S(1) to S(5). Although the FWHMs that we measure for the H$_2$ lines detected in the nuclear spectra of the QSO2s might indicate that there are no warm molecular outflows in the QSO2s, despite the presence of ionized and cold molecular outflows in all of them \citep{Ramos22,Speranza24}, further analysis of the MIRI/MRS cubes is required. J1356, J1430, and J1509 have warm molecular outflow detections based on detailed kinematic analysis of the near-infrared H$_2$ lines \citep{Ramos19,Zanchettin25}. 

Using the H$_2$ transitions from S(5) to S(1), which are the ones clearly detected in the five QSO2s, we built rotational diagrams following the methodology described in \citet{Rigopoulou02} and \citet{Pereira14}. We used the extinction corrected integrated line fluxes of the five transitions, with the corresponding errors including the fitting uncertainty reported in Table {\ref{tab:lines1}} and 10\% of flux calibration error, and the area of the region enclosing the PSF (0.4\arcsec~as radius). Since the critical densities of these H$_2$ lines are relatively low (n$_c\sim$10$^2$-10$^5$ cm$^{-3}$ at 500 K; \citealt{Bourlot99}), we can assume local thermodynamic equilibrium (LTE) conditions \citep{Roussel07}. The rotational diagrams of the QSO2s are shown in Fig. \ref{figB1} of Appendix \ref{appendixc}, which we fitted with two different models. The first is a single power law (PL) temperature distribution (dN$\sim$T$^{\beta}$dT), for which we assumed upper and lower temperatures of 3500 and 200 K, and the second is a two-temperature (2T) model including a warm (T<500 K) and a hot (T>500 K) components. The results from the fits with the PL and 2T models are summarized in Table \ref{tab:h2}.

The fits with the PL model have indices ranging from {4.3} to 5.2, which are  fully consistent with the values reported for the SINGS galaxies \citep{Togi16}, of 3.8-6.4, with an average value of 4.8$\pm$0.6 (see also \citealt{Zakamska10,Pereira14}). The power-law index can provide information
on the relative importance of gas heating by shocks, UV pumping, etc., since in that case the molecular gas is excited to higher temperatures and therefore $\beta$ becomes flatter. However, the values of $\beta$ that we find for the QSO2s are similar among them and prevent us from spotting any difference based on this. The H$_2$ masses resulting from the fit with the PL are {10.5, 12.0, 18.0, and 12.6$\times$10$^6$ M$_{\sun}$ for J1010, J1100, J1356, and J1509, and 38.2}$\times$10$^6$ M$_{\sun}$ for J1430. From the 2T model we also derive higher masses of both warm and hot molecular gas for J1430: {M$_{\rm w}$=41.0$\times$10$^6$ M$_{\sun}$ and M$_{\rm h}$=0.76$\times$10$^6$ M$_{\sun}$, whilst for the other QSO2 they are between M$_{\rm w}$=14.2-20.0$\times$10$^6$ M$_{\sun}$ and M$_{\rm h}$=0.06-0.49$\times$10$^6$ M$_{\sun}$.} The temperature of the warm gas component (T$_{\rm w}$) of the QSO2s ranges between 240 and 290 K, and that of the hot gas component (T$_{\rm h}$) is 1000 K for J1010 and J1100, {900 K for J1356}, 1100 K for J1509, and 740 K for J1430. 

The five QSO2s have molecular gas masses estimated from CO(2-1) ALMA observations \citep{Ramos22}. The highest gas masses correspond to the spiral galaxies J1100 and J1509 (2$\times$10$^{10}$ M$_{\sun}$), measured in radii of R$_{\rm CO}$=2.7 and 2.0\arcsec, respectively, which include the spiral arms. For J1010, {J1356\footnote{Here we are considering J1356N in \citet{Ramos22}, which is the member of the merger observed with JWST/MIRI and where the QSO2 is hosted.},} and J1430, the CO morphologies are more compact (R$_{\rm CO}$=0.8-1.3\arcsec) and the gas masses are lower (4-6$\times$10$^9$ M$_{\sun}$). In order to do a more accurate comparison with the warm H$_2$ masses derived here, we measured the cold molecular masses in the central kiloparsec of the QSO2s using the CO(2-1) data studied in \citet{Ramos22}, which we report in Table \ref{tab:h2}, together with the corresponding nuclear warm-to-cold gas mass ratios. Using the PL model we measure ratios of {0.005-0.017}, and using the 2T model, of {0.008-0.024}. The lowest ratios are measured for {J1356 and} J1509.

\subsection{PAH features}
\label{PAHs}

\begin{table*}
\small
\centering
\caption{PAH luminosities and equivalent widths (EWs) integrated over a local continuum and nuclear gas column density.}
\begin{tabular}{lcccccccc}
\hline
\hline
QSO2 & \multicolumn{2}{c}{PAH$_{6.2}$} & \multicolumn{3}{c}{PAH$_{11.3}$} & PAH$_{11.3/6.2}$ & PAH$_{6.2}$/S(1) & log N$_{\rm H}^{\rm CO}$ \\
     &  L$_{\rm 6.2}$ & EW$_{\rm 6.2}$ & L$_{\rm 11.3}$ & EW$_{\rm 11.3}$ & SFR$_{\rm 11.3}$  &    &   & \\
     & (10$^{\rm 42}$ erg~s$^{\rm -1}$) & (\micron) & (10$^{\rm 42}$ erg~s$^{\rm -1}$) & (\micron) & (M$_{\rm \sun}$~yr$^{\rm -1}$)  &     & & (cm$^{\rm -2}$)  \\
\hline
J1010   & {0.80$\pm$0.09}  &  0.010$\pm$0.002 & {2.70$\pm$0.36} &   0.026$\pm$0.003 & {7$\pm$1}      &    3.4$\pm${0.5}  & 8.8$\pm$0.3  & 22.3 \\
J1100   & {<0.80}          &  <0.003          & {<0.44}         &   <0.002          & {<1.0    }     &    \dots        & <8.2         & 22.5 \\
J1356   & {0.99$\pm$0.06}  &  0.016$\pm$0.003 & {1.29$\pm$0.08} &   0.037$\pm$0.002 & {3.1$\pm$0.2}  &    1.3$\pm$0.3  & 6.9$\pm$0.2  & 22.9 \\
J1430   & {1.19$\pm$0.07}  &  0.026$\pm$0.001 & {2.40$\pm$0.15} &   {0.066$\pm$0.003} & {6.0$\pm$0.4}  & {2.0$\pm$0.2}  & {5.1$\pm$0.2}  & 22.9 \\ 
J1509   & {1.60$\pm$0.08}  &  0.056$\pm$0.003 & {2.60$\pm$0.14} &   0.075$\pm$0.002 & {6.7$\pm$0.4}  &    1.6$\pm$0.2  & 13.2$\pm$0.1 & 23.0 \\ 
J1509$^{\rm IRS}$ & 7.21 & \dots & 8.02 & \dots & 21.8 & 1.11 & \dots & \dots \\
\hline
\end{tabular}
\tablefoot{Luminosities include a multiplicative factor of two to make them comparable with PAHFIT-derived measurements \citep{Smith07} and they are extinction-corrected. The last row corresponds to the measurements from the Spitzer/IRS spectrum of J1509 reported by \citet{Zakamska16}, with the luminosities multiplied by two. Upper limits at 3$\sigma$
are indicated with the < symbol for non-detections. 
SFRs listed in column 6 are calculated from L$_{\rm 11.3~\mu m}$ following Eq.~12 in \citet{2016ApJ...818...60S}, which uses a Kroupa IMF. Columns 7 and 8 correspond to the 11.3/6.2 $\mu$m PAH and 6.2 $\mu$m PAH/H$_2$0-S(1) flux ratios and they do not include the multiplicative factor of two. Column 9 is the gas column density measured from the ALMA CO(2-1) observations studied in \citet{Ramos22} using a radius of 0.4\arcsec~from the continuum peak to measure the flux, Equation 1 in \citet{Bolatto13}, and assuming R$_{\rm 21}$=1 and X$_{\rm CO}$=2$\times$10$^{20}$ cm$^{-2}$.
}
\label{tab:pahs} 
\end{table*}

We detect 6.2 and 11.3 \micron~PAH emission in {the nuclear spectra of all the QSO2s but J1100}, and also the 7.7, 8.6, and {16.5} \micron~PAH bands in J1430 and J1509 (see Fig. \ref{fig1} and Table \ref{tab:pahs}). J1100 does not show {nuclear} PAH emission, and in Table \ref{tab:pahs} we report upper limits for the 6.2 and 11.3 \micron~features. 
Using the luminosity of the 11.3 \micron~PAH feature reported in Table \ref{tab:pahs} and Eq. 12 in \citet{2016ApJ...818...60S} we measure nuclear SFRs of {7\bf $\pm$1}, {3.1$\pm$0.2}, {6.0$\pm$0.4}, and {6.7$\pm$0.4} M$_{\sun}$~yr$^{-1}$ for J1010, J1356, J1430, and J1509, and {<1.0} M$_{\sun}$~yr$^{-1}$ for J1100 (see Table \ref{tab:pahs}).  {\citet{2016ApJ...818...60S} calibrated the integrated luminosity of
the 6.2, 7.7, and 11.3 \micron~PAHs as a function of the H$_{\alpha}$-based SFR using
Spitzer/IRS data of a hundred of galaxies with infrared 
luminosities of 10$^9$-10$^{12}$ L$_{\rm \sun}$. Here we assume that all the PAH emission comes from star formation, although it has been suggested that the AGN radiation field itself could also explain the observed PAH fluxes of nearby Seyfert galaxies within the central $\sim$10-500 pc \citep{2017MNRAS.470.3071J}. On the other hand, AGN radiation can also destroy some of the PAH molecules \citep{Smith07,2010ApJ...724..140D,GarciaBernete22,GarciaBernete22a}, and therefore} here we use the 11.3 \micron~band to calculate the SFRs because it is associated with larger, neutral molecules \citep{Draine01} that should more resilient to the harsh AGN radiation and/or shocks, {following previous work \citep{Esparza18,Esparza25,2019ApJ...871..190M,GarciaBernete22a,Ramos23}.} In Table \ref{tab:pahs} we also report the PAH luminosities of J1509 measured by \citet{Zakamska16} from its Spitzer/IRS spectrum, which correspond to larger spatial scales (3.6-10.5\arcsec~$\sim$7-21 kpc). The corresponding SFR is 21.8 M$_{\sun}$~yr$^{-1}$, {almost} three times the SFR measured from the MIRI nuclear spectrum. 

The global SFRs measured from the rest-frame IR luminosities of the QSO2s obtained using far-infrared fluxes at 60 and 100 \micron~are 32, 36, {73}, 13, and 36 M$_{\sun}$~yr$^{-1}$ for J1010, J1100, J1356, J1430, and J1509 (see Table \ref{tab:sample}). These values place the QSO2s above the star formation main sequence, 
although they might contain some contribution from AGN-heated dust. \citet{Bessiere24} performed spectral synthesis modelling of the optical spectra shown in Fig. \ref{figA1}, which correspond to the central 3\arcsec~($\sim$5-6 kpc) of the QSO2s, and found SFRs of 34, 13, {1}, 3, and 3 M$_{\sun}$~yr$^{-1}$ (see Table \ref{tab:sample}). {These values are significantly lower, except in the case of J1010, than the global SFRs, already indicating that the bulk of star formation in these QSO2s comes from larger spatial scales than those probed by the SDSS fiber.} In the case of J1356, J1430, and J1509 we find good agreement between the nuclear SFRs measured from the 11.3 \micron~PAH and the spectral synthesis modelling of the QSO2s {optical} spectra, while for J1010 and J1100 the PAH-based SFRs are much lower (see Section \ref{discussion_pahs} for discussion). 

In Table \ref{tab:pahs} we report the values of the PAH 11.3/6.2 ratio for the {four} QSO2s with detections, which is an indicator of the PAH ionization fraction \citep{Galliano08,GarciaBernete22a}. We measure values of 3.4$\pm$0.5, 1.3$\pm$0.3, 2.0$\pm$0.2, and 1.6$\pm$0.2 for J1010, J1356, J1430, and J1509, which are typical of AGN-dominated regions (PAH 11.3/6.2=1-4, with an average value of $\sim$2) and higher than those reported for star-forming regions ($\lesssim$1; \citealt{GarciaBernete22,GarciaBernete24b}). 
This indicates a larger fraction of neutral PAH molecules, which produce the 11.3 \micron~feature, relative to the ionized ones  in the central kiloparsec of the QSO2s. This has been previously reported for the nuclear region of Seyfert galaxies using JWST/MIRI observations \citep{GarciaBernete22,GarciaBernete24b,Zhang24b}. Indeed, for J1509, the PAH 11.3/6.2 ratio measured from the Spitzer/IRS spectrum is 1.1 (see Table \ref{tab:pahs}), which is considerably lower than the nuclear value, of 1.6$\pm$0.2, indicating a deficit of molecules producing 6.2 \micron~PAH emission in the nucleus of this QSO2.

\section{Discussion}
\label{discussion}

\subsection{Silicate features and gas column densities}
\label{discussion_silicates}

One of the most striking results derived just from visual inspection of the mid-infrared spectra shown in Fig. \ref{fig1} are the differences in the continuum shape and silicate feature strength of the five QSO2s. The brightest QSO2 in the mid-infrared is J1100, followed by J1010, J1430, J1356, and J1509 (see caption of Fig. \ref{fig1}). 
This is consistent with the observed K-band excesses reported for J1010 and J1100 \citep{Shangguan19,Jarvis20}, indicative of an important contribution from AGN-heated dust \citep{Mor09,Ramos09b,Ramos11,GarciaBernete17}. In the framework of clumpy torus models, the silicate features (either in absorption or emission) are always relatively weak, unlike in smooth-density distributions, because both illuminated and shadowed cloud sides contribute to them. While most views of the torus in type-2 AGN intercept absorbing shadowed cloud faces, silicate emission from some bright cloud faces fills in the feature, making it shallower, absent, or in emission \citep{nenkova08}. The latter is likely the case of J1010 and J1100, which show 9.7 \micron~silicate features in emission (S$_{\rm 9.7}$={0.5 and 0.1}), as well as the 18 and 23 \micron~features (see Table \ref{tab1}), despite their type-2 AGN classification in the optical. 

However, by looking at the optical spectra of the QSO2s shown in Fig.~\ref{figA1} it can be noticed that the H$_{\alpha}$+[NII] profile of J1010 appears broader than those of the other QSO2s and H$_{\beta}$ {(see Fig. \ref{figA1bis})}, which could be consistent with a type 1.9 AGN classification\footnote{Having an intermediate view of the AGN where broad emission from the broad-line region (BLR) is observable in H$_{\alpha}$ but not in H$_{\beta}$.} that would also explain its silicate emission feature. For reference, \citet{Mariela20} reported an average value of S$_{\rm 9.7}$=0.11$\pm^{0.15}_{0.36}$ (peak wavelength at 10.3$\pm^{0.7}_{0.9}$ \micron) for a sample of 67 type-1 AGN at z<1 with Spitzer/IRS spectra. For the 18 \micron~feature, they found S$_{\rm 18}$=0.14$\pm$0.06 (peak wavelength at 17.3$\pm^{0.4}_{0.7}$ \micron). These values are similar to those measured for J1100 (see Table \ref{tab1}), whilst the strength of the 9.7 \micron~feature of J1010 is among the 10 highest values measured for the type-1 AGN in \citet{Mariela20}, which range between 0.3 and 0.5. 

To test the possibility {that J1010 might be a type 1.9 AGN} we performed a multi-component Gaussian fit of the H$_{\beta}$, [OIII], and H$_{\alpha}$+[NII] profiles {detected in the continuum-subtracted SDSS spectrum of the QSO2} (see Fig. \ref{figA2} in Appendix \ref{appendix}). From this analysis we find an extra broad H$_{\alpha}$ component of {FWHM=3400$\pm$100 km~s$^{-1}$ and blueshifted by 1000$\pm$70 km~s$^{-1}$} relative to systemic, which could be associated either with the BLR (i.e. type 1.9 AGN) or with a weak, additional outflow component (the five QSO2s have ionized outflows detected in [OIII]; \citealt{Bessiere24,Speranza24}) that we do not see in H$_{\beta}$ and [OIII]. The type 1.9 AGN scenario is unlikely because the blueshift that we measure is higher than expected for the BLR at the bolometric luminosities of the QSO2s and even higher ({$\sim$50-500 km~s$^{-1}$;} \citealt{deconto23,deconto24}). Therefore, based on the analysis of the SDSS spectrum of J1010, we favor the outflow scenario to explain the extra broad component of H$_{\alpha}$ in J1010.

J1356, J1430, and J1509, on the other hand, do not show K-band excesses and they have silicate features in absorption (S$_{9.7}$=-1.0, -0.3, and -0.2, respectively), which are more typical of type-2 AGN, and of QSO2s in particular (see Table \ref{tab1} and Figs. \ref{fig2} and \ref{fig3}). \citet{Zakamska16} reported a mean value of S$_{9.7}$=-0.41$\pm$0.48 (median of -0.30) for a sample of 46 type-2 AGN (including QSO2s and Seyferts) at z=[0.04,0.7] observed with Spitzer/IRS (see Fig. \ref{fig3}). Only one of the targets in their sample shows the silicate feature in emission, with S$_{9.7}$=0.27. \citet{Sturm06} reported tentative evidence of 9.7 \micron~silicate in weak emission or absent in the spectra of five obscured quasars with redshifts between 0.2 and 1.4, but based on Spitzer/IRS spectra covering a narrow spectral range, of between $\sim$4-6 \micron~for the bluest wavelength of the spectra and $\sim$11-14 \micron~for the reddest, and without subtracting PAH emission.

In the recent work of \citet{GarciaBernete24}, the nuclear mid-infrared spectra of the six Seyfert 2 galaxies from GATOS observed with JWST/MIRI, which have X-ray column densities of log N$_H$=22.2-24.3 cm$^{-2}$, show the 9.7 and 18 \micron~silicate features in absorption, with S$_{9.7}$=[-2.2,-0.3] and S$_{18}$=[-0.5,0.0]. No X-ray column density values are reported in the literature for our targets, except for J1430, of log N$_H\sim$23.7 cm$^{-2}$ \citep{2018ApJ...856L...1L}. This is the expected column density for a QSO2 \citep{Zakamska08,Zakamska16}, and consistent with a silicate feature seen in shallow absorption \citep{Shi06}. Here we estimated the nuclear gas column densities of the five QSO2s from the CO(2-1) ALMA observations used in \citet{Ramos22}, by integrating the flux in an aperture of 0.4\arcsec~radius from the continuum peak and using Eq. 1 in \citet{Bolatto13}. The gas column density values are reported in Table \ref{tab:pahs} and they range between 22.3 and 23.0 cm$^{-2}$. These are modest nuclear gas column densities considering that the targets are obscured QSO2s, although the relatively low extinctions measured from the H$_{\alpha}$/H$_{\beta}$ ratios already pointed in that direction (see Table \ref{tab:sample}). 

The lowest CO column densities are those of J1010 and J1100: the QSO2s with the 9.7 \micron~silicate feature in emission and with weak and no nuclear PAH emission, respectively (see Section \ref{discussion_pahs}). Furthermore, in {these two} QSO2s we detect the 18 and 23 \micron~silicate features {in emission}, which is expected in the case of modest obscuration \citep{Spoon22}. J1430 is an intermediate case, where the 9.7 \micron~silicate feature is in absorption, but the 18 and 23 \micron~silicate features are in emission. 
Only in {J1356 and J1509, the most obscured QSO2s in the sample together with J1430 (log N$_{\rm H}^{\rm CO}$=23 cm$^{-3}$),} we do not detect the 23 \micron~silicate features, {the 18 \micron~feature is in weak emission (S$_{18}$=0.06), and they both show the {\color {black} ro-vibrational 4.67 \micron~CO band, several \micron~H$_2$O lines from $\sim$5.5 to 6.2 \micron, the 6 \micron~water ice band, and the 6.85 and 7.25 \micron~aliphatic grains in absorption} (see Figs. \ref{fig1} and \ref{fig:ices}). For the water ice, we measure strengths of S$_6$=-0.4 and -0.2 for J1356 and J1509, which are similar to the values reported in \citet{GarciaBernete24} for the six GATOS Seyfert 2 galaxies (S$_6$=[-0.5,0.0]).} This would be in agreement with \citet{GarciaBernete24}, who reported a correlation between the depth of the water ice and the gas column density, albeit derived from the X-rays in their case. However, in the case of J1430, for which we also measure log N$_{\rm H}^{\rm CO}$=23 cm$^{-3}$, we do not detect any of the ices.

The X-ray column density of J1430 (log N$_{\rm H}^{\rm X}$=23.7 cm$^{-2}$) is higher than the one estimated from CO, which is not surprising if we consider that {X-ray obscuration comes not only from the CO-emitting gas producing N$_{\rm H}^{\rm CO}$ but also from dust-free gas within the sublimation region \citep{Ramos17review}}. Indeed, \citet{2021arXiv210410227G} compared X-ray and CO-based column densities measured for the torus of 14 AGN from the GATOS and NUclei of GAlaxies (NUGA) samples and found that while galaxies with log N$_{\rm H}^{\rm X}$<22.0 cm$^{-2}$ show N$_{\rm H}^{\rm CO}$ values above the 1:1 relation, those with log N$_{\rm H}^{\rm X}$>22.0 cm$^{-2}$ have N$_{\rm H}^{\rm CO}$ lying closer or below the 1:1 line, as it is the case of J1430.  

In summary, the relatively low gas column densities that we measure for the nuclear region of J1010 and J1100, together with their strong mid-infrared continua, explain the detection of their silicate features in emission {and the non-detection of {\color {black} absorption} bands. On the other hand, J1356, J1430, and J1509 have higher gas column densities derived from CO observations and therefore show the 9.7 \micron~silicate feature in absorption, but these column densities are not high enough to produce 18 and 23 \micron~silicate bands in absorption, as it is the case for ULIRGs \citep{Spoon06,Spoon22,Donnan23}. Nevertheless, in J1356 and J1509 
the 23 \micron~silicate features appear flat and we detect {\color {black} several absorption} bands, indicating higher nuclear obscuration than in J1430. X-ray observations of these QSO2s would be required to confirm this.}

\subsection{Gas excitation mechanisms}
\label{discussion:excitation}

Fig. \ref{models} shows that the [NeIII]/[NeII] and [NeV]/[NeII] ratios of J1100, J1356, J1430, and J1509 are similar to those of the GATOS Seyfert galaxies observed with JWST/MIRI \citep{Hermosa24,Zhang24a} and they can be successfully reproduced with the AGN photoionization models from \citet{Feltre16,Feltre23}. Instead, J1010 occupies a different region of the diagram, which can be explained by AGN photoionization with a low ionization parameter (log U$\sim$-3), but also by AGN+shocks models with low velocities (v$\sim$200-300 km~s$^{-1}$) and with AGN+star formation models with low ionization parameter (log U$\sim$-2.5), as shown in Fig. \ref{models} \citep{Feltre23}. From this comparison with different models it is clear that the ionizing continuum of J1010 is different from those of the other four QSO2s. This was already hinted by the HeII$\lambda$4686/H$_{\beta}$ ratios measured from the optical SDSS spectrum shown in Fig. \ref{figA1}. While for J1010 we measure a remarkably low ratio of 0.0$\pm$0.4, the other four QSO2s show typical AGN values \citep{Oh17}, ranging from 0.1$\pm$0.1 in J1430, to 0.2-0.3 in the other three.

\citet{Ardila11} reported near-infrared spectra of 54 nearby Seyfert galaxies and found coronal emission lines with IPs between 125 and 450 eV detected at 3$\sigma$ in 67\% (36 AGN) of the sample. They claimed that, although most of the non-detections are due to either not enough spatial resolution or increasing object distance, there are AGN where the lack of coronal emission may be genuine, possibly due to a non-standard ionizing continuum such as a hard X-ray spectrum lacking photons below a few keV (see e.g. \citealt{Alonso24} for recent JWST/MIRI observations of the X-ray weak quasar Mrk231). This is because coronal emission becomes stronger with increasing X-ray emission and steeper X-ray photon index (i.e., softer X-ray spectra). X-ray observations of the QSO2s studied here can therefore contribute to understand the diverse coronal line emission that we observe in their mid-infrared spectra.

Shocks induced by jets have been also invoked to explain the presence of coronal line emission on scales from hundreds to thousands of parsecs from the AGN \citep{Tadhunter87,Tadhunter88,Worrall12,Ramos17,Ardila17,Ardila20}, as well as enhanced warm H$_2$ emission \citep{Pereira22,Davies24,Riffel25}. PAH/H$_2$ ratios are usually low in galaxies with strong jet-ISM interactions because shocks are one of the main excitation mechanisms of H$_2$, making PAH/H$_2$ lower than in galaxies with similar PAH luminosities but without jet-ISM interactions \citep{Ogle10,Nesvadba10,GarciaBernete24b,Riffel25}. Indeed, in the case of Teacup (J1430), where we know that there is strong jet-ISM interaction in the central kiloparsec (see below), the value of 6.2 \micron~PAH/S(1) is one of the lowest, of 5.1$\pm$0.2 (see Table \ref{tab:pahs}). For J1100 we measure a ratio of <8.2, albeit is due to the lack of PAH emission rather than to enhanced S(1), and for J1010 and J1356, of 6.9$\pm$0.2 and 8.8$\pm$0.3 respectively. For reference, \citet{GarciaBernete24b} reported values of 6.2 \micron~PAH/H$_2$ S(1)>12 for star-forming regions and of <5 for AGN-dominated regions. Considering this, the value that we measure for J1509, of 13.2{$\pm$0.2}, indicates a low contribution from shocks in the presence of star formation.

Considering the similar cold molecular masses measured in the central kiloparsec of the QSO2s, of 1-2$\times$10$^{9}$ M$_{\odot}$, the larger mass of warm molecular gas (both M$_{\rm w}$ and M$_{\rm h}$ in Table \ref{tab:h2}) that we find in J1430 {compared to the other QSO2s indicates higher} H$_2$ excitation in the central kiloparsec of {this galaxy. Indeed, J1430} is the only one of the five studied here with a spatially-resolved jet-like structure in VLA observations at 0.2\arcsec~resolution \citep{Jarvis19}. This structure is a low-power (P$_{\rm jet}\sim$10$^{43}$ erg~s$^{-1}$) compact jet of $\sim$1 kpc size, that is almost coplanar with the molecular gas disc \citep{Audibert23}. Using ALMA CO(2-1) and CO(3-2) observations of this target, \citet{Audibert23} found higher brightness temperature ratios (T$_{32}$/T$_{21}$) and CO velocity dispersion in the direction perpendicular to the jet (see e.g. \citealt{Venturi23} for examples of similar behavior in the case of the ionized gas), suggesting that the jet is compressing the molecular gas, inducing shocks, and enhancing turbulence there. This has been predicted by hydrodynamical simulations of jet-gas interactions \citep{Wagner11,Mukherjee18}, including one tailored to this QSO2 \citep{Meenakshi22,Audibert23}. Therefore, the jet-ISM interaction in J1430 is likely shock-exciting the H$_2$ emitting gas to high energy levels, resulting in the higher rotational line fluxes and corresponding masses of warm molecular gas that we measure here. A comparison between the spatially resolved properties of the cold and warm molecular gas will be the subject of a forthcoming work. 

On the other hand, we find the highest temperatures of the {hot} molecular gas component {(T$_{\rm h}$)} in J1509, the QSO2 with the strongest H$_2$ emission probing gas at T>500 K but with the least contribution from shocks according to the 6.2 \micron~PAH/S(1) ratio. In the case of this QSO2, the intense emission of mid-infrared H$_2$ emitting gas at high temperatures might be due to UV heating and/or turbulence \citep{Kristensen23}. Indeed, this QSO2 has a multi-phase outflow with PA$\sim$-40$^{\rm o}$, studied using low- and high-ionization atomic lines \citep{Ramos19,Speranza24}, near-infrared H$_2$ lines \citep{Ramos19}, and CO \citep{Ramos22}. The atomic lines detected in the mid-infrared nuclear spectrum of J1509 show the blueshifted wing characteristic of the ionized outflow, but not the rotational H$_2$ lines (see Fig. \ref{fig1} and Table \ref{tab:kin}). 

Using the warm molecular gas masses and the total gas masses measured from CO in the central kiloparsec of the QSO2s, both reported in Table \ref{tab:h2}, we find warm-to-cold gas mass ratios of {$\sim$1-2\%}. These ratios are among the lowest measured from ISO observations of nearby Seyfert galaxies ($\sim$2-35\%; \citealt{Rigopoulou02}), {and from Spitzer/IRS spectra of nearby Seyferts in LIRGs (1-20\%; \citealt{Pereira14}), and of non-active galaxies in the SINGs sample, between 1 and >30\% \citep{Roussel07}.}

\subsection{Electron density}
\label{discussion:density}

The ratio of [NeV]$_{\rm 14.3/24.3}$ (see Table \ref{tab:density}) can be used to estimate the electron density of the gas in the nuclear region of the QSO2s. The values that we measure for the QSO2s are more similar to those derived from the optical trans-auroral lines than from the [SII]$\lambda\lambda$6716,6731 \AA~doublet. This is because the trans-auroral lines are sensitive to a wider range of electron densities, supporting their use as density indicators for the nuclear regions of AGN, including outflows \citep{Holt11,Rose18,Ramos19,Davies20,Speranza22,Speranza24,Holden23a,Holden23b,Bessiere24}. 

The [NeV]$_{\rm 14.3/24.3}$ ratio is >2.89 for J1010, since we do not detect [NeV]$_{\rm 24.3}$ in the spectrum shown in Fig. \ref{fig1}, and this is consistent with having a high electron density, of log n$_e$>4.07 cm$^{-3}$. Interestingly, the value that we measure from the trans-auroral lines, of log n$_e$=4.58$\pm$0.03 cm$^{-3}$, is higher than the critical density of [NeV]$_{\rm 24.3}$ (log n$_c$=4.42 cm$^{-3}$) and close to that of [NeV]$_{\rm 14.3}$ (log n$_c$=4.68 cm$^{-3}$). This might be contributing to the non-detection of [NeV]$_{\rm 24.3}$ in the nuclear spectrum of J1010, assuming that the density derived from the trans-auroral line ratio measured from the SDSS spectrum (central 3\arcsec$\sim$5 kpc) of the QSO2 is equal or lower than the nuclear density, for which we have a lower limit of log n$_e$>4.07 cm$^{-3}$. Other coronal lines such as [MgVII]$_{\rm 5.50}$, [MgV]$_{\rm 5.61}$, and [NeVI]$_{\rm 7.65}$ have higher critical densities, of log n$_c$=6-7 cm$^{-3}$, and they are detected in the nuclear spectrum of J1010, albeit they are weak. Therefore, in the case of J1010, {an X-ray spectrum lacking a strong soft component might explain} the weak coronal line emission in comparison with the other QSO2s. Moreover, in the particular case of [NeV]$_{\rm 24.3}$, the high density inferred from the [NeV]$_{\rm 14.3/24.3}$ ratio and from the optical trans-auroral lines might be further suppressing its emission relative to that of [NeV]$_{\rm 14.3}$ within the nuclear region. Observations of a larger number of QSO2s probing a range of electron densities, such as the QSOFEED sample (log n$_e$=2.3-4.6; \citealt{Bessiere24}), and X-ray photon indices are required to confirm these tantalizing trends.

\subsection{Nuclear PAH emission}
\label{discussion_pahs}

Using the luminosity of the 11.3 \micron~feature and {assuming that it comes from star formation only,} we derived nuclear SFRs for the QSO2s (see Table \ref{tab:pahs}). The values that we measure for J1356, J1430, and J1509, of {3.1$\pm$0.2}, {6.0$\pm$0.4,} and {6.7$\pm$0.4} M$_{\sun}$~yr$^{-1}$, are in good agreement with those measured from spectral synthesis modelling of the optical SDSS spectra shown in Fig. \ref{figA1}, of {1, 3, and 3 ($\pm$1)} M$_{\sun}$~yr$^{-1}$ respectively \citep{Bessiere24}. In the case of J1010 and J1100, however, the PAH-derived SFRs, of {7$\pm$1 and <1.0} M$_{\sun}$~yr$^{-1}$, are significantly lower than those reported by \citet{Bessiere24} for the central 5-6 kpc of these QSO2s, of {34$\pm$8} and {13$\pm$3} M$_{\sun}$~yr$^{-1}$. This discrepancy can be accounted for by i) reduced star formation in the central kiloparsec of these QSO2s (i.e. negative feedback; see e.g. \citealt{Esparza18,Bessiere22}); ii) dilution by their stronger mid-infrared continuum (see e.g. \citealt{Alonso14,Ramos14}); and/or {iii)} destruction of PAH molecules due to the combination of strong AGN radiation and relatively low gas column densities (see Table \ref{tab:pahs} and \citealt{Alonso20}). 

Although it is clear that there is some level of dilution from the lower EWs that we measure for both the 6.2 and 11.3 \micron~features in J1010 and J1100 as compared to J1430 and J1509 (see Table \ref{tab:pahs}), {with J1356 lying between them,} the integrated PAH flux should not be affected \citep{Alonso14,Esparza18}, and therefore we should be able to measure higher SFRs if present \citep{Ramos23}. Attending to their ionized and cold molecular outflow properties, J1010 and J1100 do not appear different from J1430 and J1509 \citep{Ramos22,Bessiere24,Speranza24}. Therefore, in principle we do not expect more efficient negative feedback in the nuclei of J1010 and J1100 that could support the first scenario (i). Considering this, here we favor the third scenario (iii) for explaining the low PAH-based SFRs measured for J1010 and J1100, which have modest gas column densities (see Table \ref{tab:pahs}). The higher PAH$_{\rm 11.3/6.2}$ ratio that we find in J1010, {of 3.4$\pm$0.5}, compared with those of J1356, J1430 and J1509, which range between 1.3 and 2.0, further supports the destruction of some PAH molecules in the nuclear region of this QSO2. In harsh environments, this can be explained as preferential destruction of ionised PAH molecules via Coulomb explosion \citep{GarciaBernete24b}. It is worth mentioning, however, that another possibility for explaining the dearth of ionised PAHs is due to high recombination rates in the nuclear region of AGN \citep{Rigopoulou24}. The forthcoming integral field analysis of the MIRI/MRS cubes will permit us to study the spatial distribution of the PAHs and the emission line features.

\section{Conclusions}

Here we presented new mid-infrared nuclear spectra of the central 0.7-1.3 kpc of five optically-selected QSO2s at redshift z$\sim$0.1 from the QSOFEED sample, obtained with the MRS module of the JWST instrument MIRI. These QSO2s, {whose [OIII] and CO(2-1) emission has been studied in detail in \citet{Ramos22} and \citet{Speranza24} respectively,} have bolometric luminosities of log L$_{\rm bol}$=45.5 to 46 erg~s$^{-1}$, global SFRs of 13-73 M$_{\sun}$~yr$^{-1}$, and practically identical optical spectra in terms of spectral shape and [OIII] luminosities. Regardless of this, their nuclear mid-infrared spectra show an unexpected diversity that we summarize below. 

\begin{itemize}

    \item The 9.7 \micron~silicate feature appears in emission in J1010 (S$_{9.7}$=0.5), in weak emission in J1100 (S$_{9.7}$=0.1), in shallow absorption in J1430 and J1509 (S$_{9.7}$=-0.3 and -0.2 respectively), {and in relatively deep absorption in J1356 (S$_{9.7}$=-1.0).} We also detect {both} the 18 $\mu$m silicate feature and the 23 \micron~band of crystalline silicates in emission in J1010, J1100, and J1430. The strengths of these features indicate moderate obscuration in these QSO2s. 
    
    \item In J1356 and J1509, the QSO2 with higher column densities in the sample together with J1430 (log N$_H^{\rm CO}$=23 cm$^{-2}$), the 23 \micron~silicate band is flat, the 18 \micron~silicate feature is in weak emission, and we {\color {black} detect 4.67 \micron~CO, 5.5-6.2 \micron~H$_2$O, 6 \micron~water ice, and 6.85 and 7.25 \micron~aliphatic grain bands in absorption}. These features are indicative of higher nuclear obscuration in these two QSO2s.

    \item The profiles of the atomic lines can be reproduced with two Gaussian components (three in the case of J1356), with the broadest one likely associated with outflowing gas (FWHMs$\sim$900-1600 km~s$^{-1}$). The broad component of the [NeV] line is broader than that of [NeII] in all the QSO2s but J1356, which might be indicating that [NeV] is probing outflowing gas closer to the AGN. The rotational H$_2$ lines can be fitted with a single Gaussian of FWHM$\sim$300-600 km~s$^{-1}$. 
   
    \item We measure [NeV]/[NeII] ratios ranging from 0.1 to 2.1 and [NeIII]/[NeII] from 1.0 to 3.5, indicating different coronal line and ionizing continuum strengths. J1100 shows the largest ratios, indicative of an energetic ionizing continuum and strong coronal line emission, J1356, J1430 and J1509 have typical values of AGN, including QSO2s and Seyferts, and J1010 shows the lowest ratios, which might be indicative of a non-standard ionizing continuum in this QSO2. 

    \item Using the [NeV]$_{\rm 14.3/24.3}$ ratio we estimate the electron density of the coronal gas. The values that we measure, of log n$_e$>2.8 cm$^{-3}$, are closer to the electron densities estimated from the optical trans-auroral lines than from the [SII]$\lambda\lambda$6716,6731 \AA~doublet. For J1010 we do not detect the [NeV]$_{\rm 24.3}$ \micron~line in the nuclear spectrum, which sets a lower limit for the density of log n$_e$>4.07 cm$^{-3}$. 
    
    \item The value of n$_e$ measured in the central 5 kpc of J1010 from the optical trans-auroral lines detected in the SDSS spectrum (log n$_e$=4.58$\pm$0.03 cm$^{-3}$) is higher than the critical density of [NeV]$_{\rm 24.3}$. This high density might be contributing to suppress its emission relative to that of [NeV]$_{\rm 14.3}$ in the central kiloparsec of this QSO2 if the density there is equal or higher than the value measured from the SDSS spectrum. 
    
    \item We detect the rotational H$_2$ lines from S(9) to S(1) in J1010, J1356, J1430, and J1509. In J1100 the lines from S(6) to S(9) are rather weak, possibly because of its strong continuum. Using rotational diagrams involving the S(5) to S(1) transitions, we measure nuclear warm molecular gas masses of 1-4$\times$10$^7$ M$_{\sun}$, with molecular gas excitation most likely due to jet-induced shocks and turbulence in the case of J1430, and to UV heating and/or turbulence in J1509. 

    \item Using CO-derived molecular gas masses measured in the central kiloparsec of the QSO2s ({1-2$\times$10$^9$ M$_{\sun}$)} we obtain warm-to-cold gas mass ratios of 1-2\%, which are among the lowest reported for nearby Seyfert galaxies using either ISO or Spitzer observations (1-35\%).

    \item We detect 6.2 and 11.3 \micron~PAH emission in the central kiloparsec of all the QSO2s but J1100, as well as the 7.7, 8.6, and 16.5 \micron~PAH bands in J1430 and J1509. The PAH 11.3/6.2 ratios range between 1.3 and 3.4, indicating a larger fraction of the neutral PAH molecules producing the 11.3 \micron~feature in the nuclear region of these QSO2s.

    \item Using the 11.3 \micron~PAH luminosity, we obtain SFRs of $\sim${3-7} M$_{\sun}$~yr$^{-1}$ for the QSO2s with detections and <1 M$_{\sun}$~yr$^{-1}$ for J1100. In the case of J1010 and J1100, these values are significantly lower than the SFRs measured from spectral synthesis modelling of optical spectra probing the central 5-6 kpc of the QSO2s. This is likely due to the destruction of some PAH molecules by the AGN radiation in the central kiloparsec of the galaxies.

\end{itemize}

The high angular and spectral resolution of JWST/MIRI, coupled with the unprecedented sensitivity that it offers in the mid-infrared, have revealed an unexpected diversity of nuclear spectral features in this small sample of QSO2s. In this work we have started exploring the role of various AGN and galaxy properties on explaining some of the spectral differences listed above, but larger samples are now required to fully understand the diversity of QSO2s' mid-infrared nuclear spectra. 

\begin{acknowledgements}
The authors acknowledge the anonymous referee for their constructive report. CRA acknowledges the hospitality of the Kavli Institute for Cosmology of the University of Cambridge, where a substantial part of this manuscript was written, in August 2024. This stay was funded by the EU H2020-MSCA-ITN-2019 Project 860744 “BiD4BESt: Big Data applications for black hole Evolution STudies.” CRA also thanks Anna Feltre for the use of her models, and Ignacio Martín Navarro and Andrés Asensio Ramos for useful discussions. 
CRA, AA, JAP, and PC acknowledge support from the Agencia Estatal de Investigaci\'on of the Ministerio de Ciencia, Innovaci\'on y Universidades (MCIU/AEI) under the grant ``Tracking active galactic nuclei feedback from parsec to kiloparsec scales'', with reference PID2022$-$141105NB$-$I00 and the European Regional Development Fund (ERDF). IGB is supported by the Programa Atracci\'on de Talento Investigador ``C\'esar Nombela'' via grant 2023-T1/TEC-29030 funded by the Community of Madrid. MPS acknowledges support under grants RYC2021-033094-I, CNS2023-145506 and PID2023-146667NB-I00 funded by MCIN/AEI/10.13039/501100011033 and the European Union NextGenerationEU/PRTR. RM acknowledges support by the Science and Technology Facilities Council (STFC), by the ERC through Advanced Grant 695671 ”QUENCH”, and by the UKRI Frontier Research grant RISEandFALL. RM also acknowledges funding from a research professorship from the Royal Society. AAH acknowledges support from grant PID2021-124665NB-I00 funded by the Spanish Ministry of Science and Innovation and the State Agency of Research MCIN/AEI/10.13039/501100011033 and ERDF A way of making Europe. SGB acknowledges support from the Spanish grant PID2022-138560NB-I00, funded by MCIN/AEI/10.13039/501100011033/FEDER, EU. OGM acknowledges financial support from PAPIIT UNAM project IN109123 and Ciencia de Frontera CONAHCyT project CF-2023-G100. DR acknowledges support from STFC through grants ST/S000488/1 and ST/W000903/1.
\end{acknowledgements}

\bibliographystyle{aa} 
\bibliography{aa53549-24} 

\begin{appendix} 
\onecolumn
\section{Optical spectra of the QSO2s}\label{appendix}

\begin{figure}[h!]
   \centering
  \includegraphics[width=0.9\columnwidth]{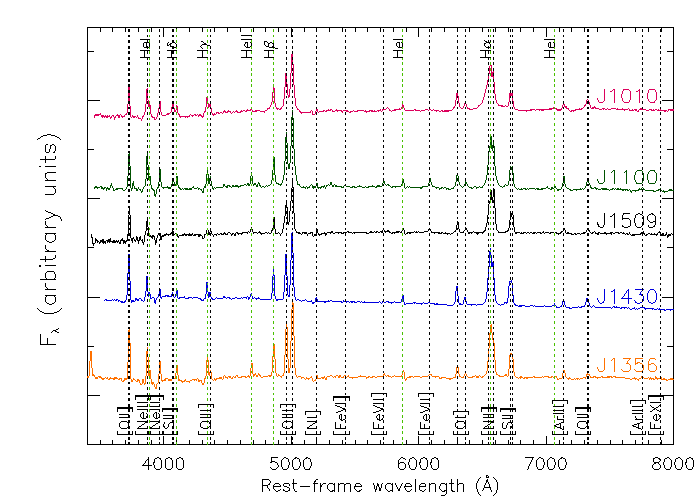}
   \caption{SDSS optical spectra of the central 3\arcsec~of the QSO2s ($\sim$5-6 kpc at z=0.09-0.12). The spectra have been scaled in the Y-axis using a multiplicative factor to better show the spectral features and the color code is the same as in Fig. \ref{fig1}. The most intense emission line features are labelled and their wavelengths are indicated with dotted lines: green for the permitted lines and black from the forbidden ones.}
              \label{figA1}%
    \end{figure}

\begin{figure}[h!]
   \centering
  {\par\includegraphics[width=0.45\columnwidth]{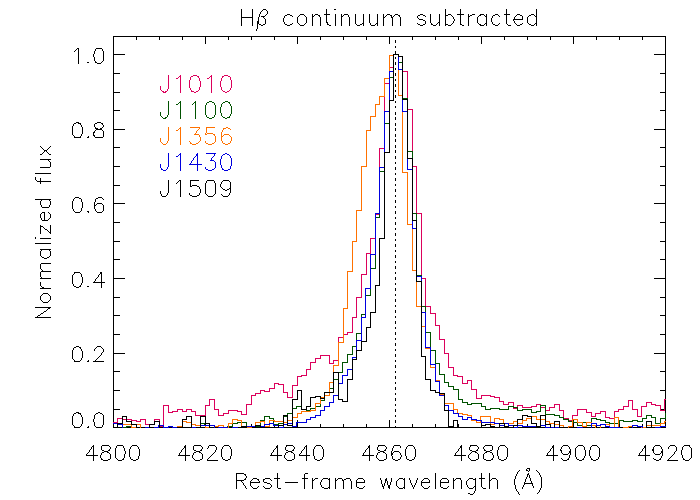}
   \includegraphics[width=0.45\columnwidth]{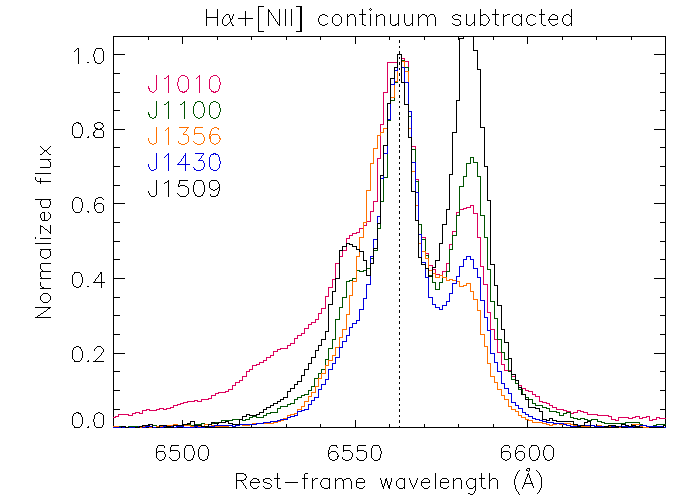}\par}
   \caption{Continuum-subtracted spectrum of J1010 around the H$_{\beta}$ (left panel) H$_{\alpha}$+[NII] emission line complex (right panel), both normalized to the peak of the recombination line, respectively. The vertical dotted lines correspond to the rest-frame wavelength of H$_{\beta}$ and H$_{\alpha}$.}
              \label{figA1bis}%
    \end{figure}

In order to look for potential differences between the optical spectra of the five QSO2s studied here, we downloaded their spectra from the Sloan Digital Sky Survey (SDSS) archive (see Fig.~\ref{figA1}). They correspond to the Legacy survey \citep{abazajian09}, which used a 3\arcsec~diameter fibre and covered an observed wavelength range of 3800 -- 9200 \AA\ ($\sim$3455-8360 \AA\ at the average redshift of the QSOFEED sample, z=0.11) with a spectral resolution of R=1800--2200. The same spectra were used in \citet{Bessiere24} to study the underlying stellar populations and the [OIII] kinematics using a non-parametric analysis. No significant differences are seen among the optical spectra of the QSO2s in terms of spectral shape and emission line kinematics, except for the broader H$_{\alpha}$+[NII] profile of J1010 (see Fig. \ref{figA1bis}). J1100 shows stronger coronal emission than the other QSO2s (e.g. [FeVII]$\lambda$6087 \AA).

Our QSO2s are part of \citet{2008AJ....136.2373R} optically-selected sample of 887 objects with z<0.83, from where type-1 AGN were removed. To do so, the latter authors performed a non-parametric fitting of the [OIII] lines and used these profiles to fit the H$_{\alpha}$+[NII] complex. When no significant residuals were left the AGN was considered a narrow-line (i.e., type-2) AGN and therefore retained in the sample. Otherwise the object was classified a broad-line (type-1) AGN and removed from it. An example of this procedure can be found in Fig. 1 of \citet{2008AJ....136.2373R}, where the example of confirmed QSO2 is one of our targets, J1100. 

\begin{figure}[h!]
   \centering
  {\par\includegraphics[width=0.45\columnwidth]{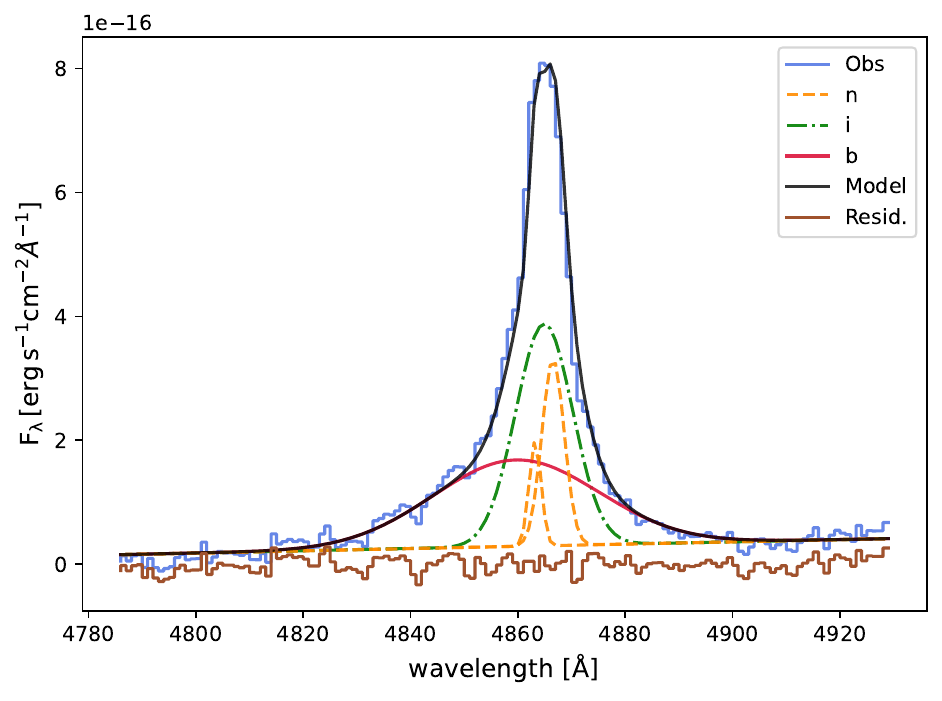}
  \includegraphics[width=0.45\columnwidth]{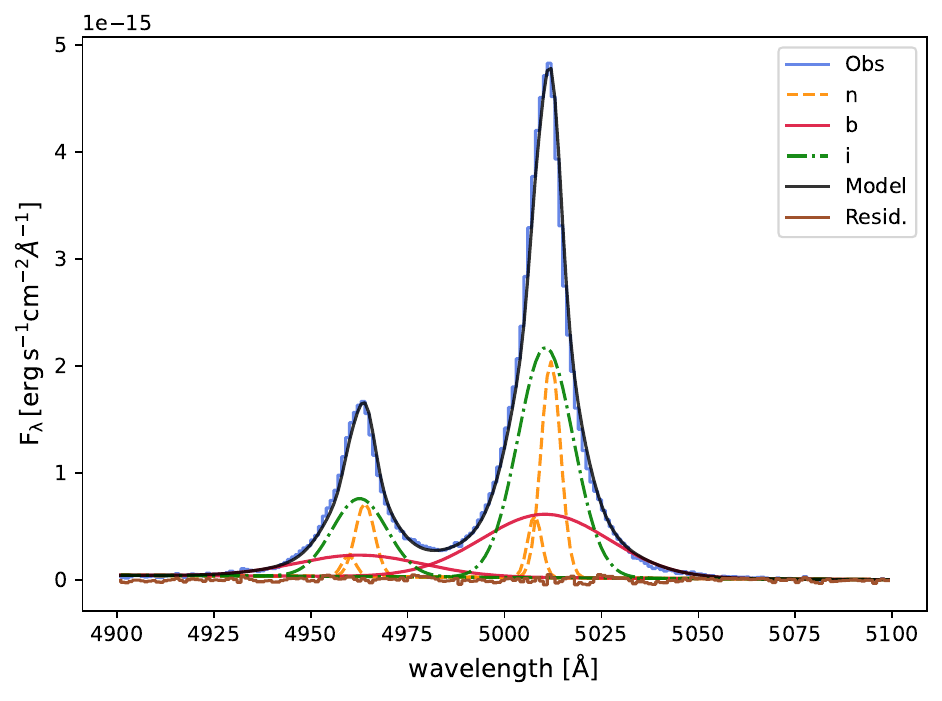}
  \includegraphics[width=0.45\columnwidth]{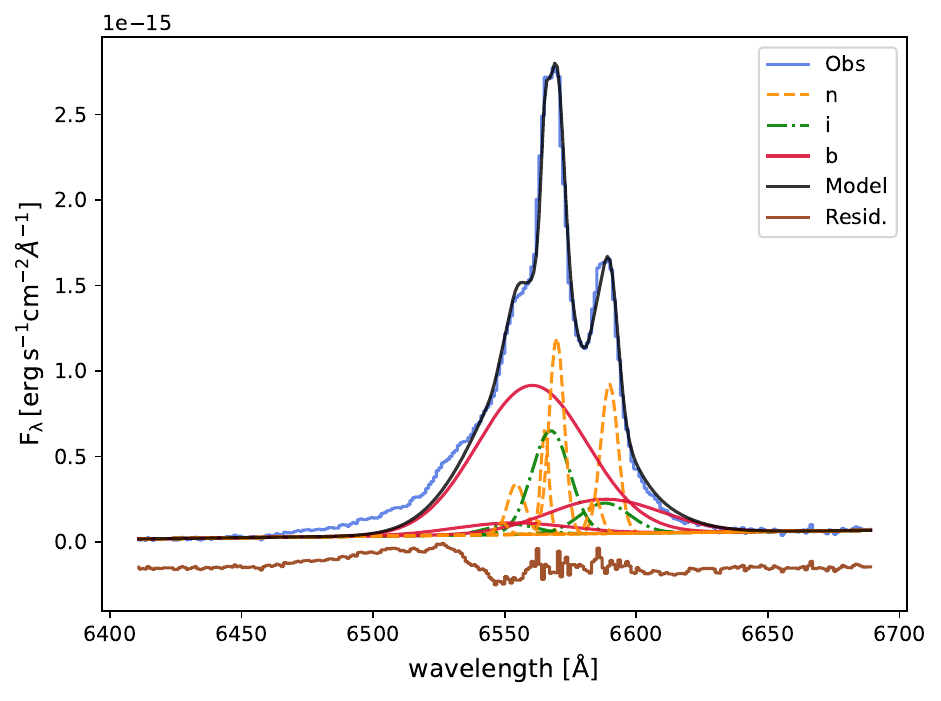}
  \includegraphics[width=0.45\columnwidth]{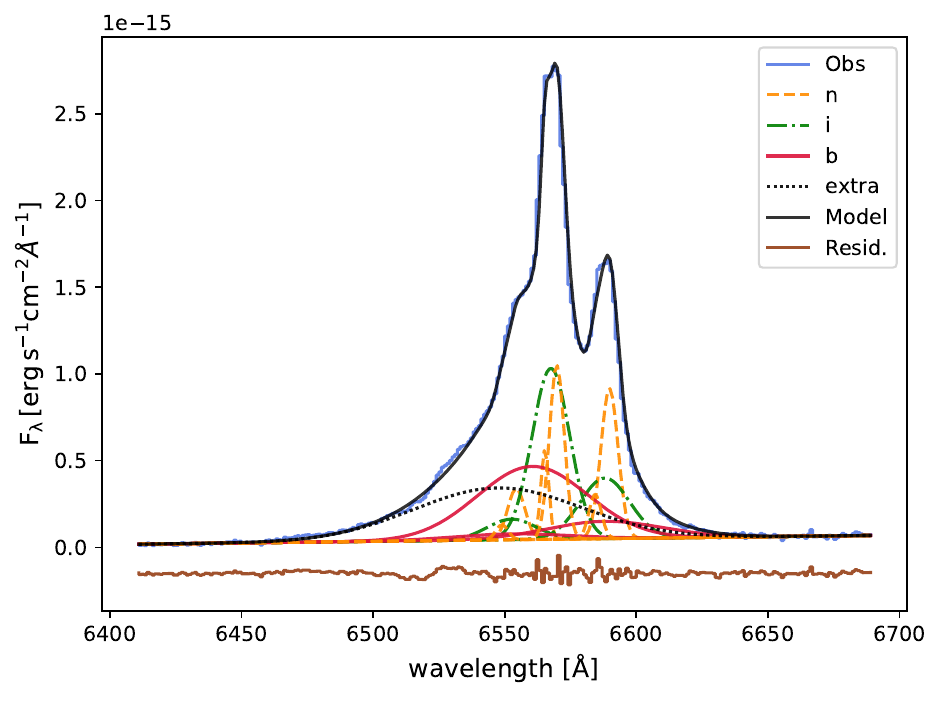}\par}
   \caption{Fits of the H$_{\beta}$, [OIII], and H$_{\alpha}$+[NII] profiles detected in the continuum-subtracted SDSS spectrum of J1010. The observed spectra are shown in blue, the model including all the fitted components in black, and the residuals in brown. Same colors correspond to the same physical components: narrow (n; orange dashed), intermediate (i; green dot-dashed), broad (b; red solid), and extra (black dotted).}
              \label{figA2}
    \end{figure}

In the case of J1010, to test its possible classification as a type-1.9 AGN \citep{Osterbrock81}, we took the continuum-subtracted SDSS spectrum from \citet{Bessiere24} and performed a multi-component Gaussian fit of the H$_{\beta}$ and [OIII] profiles, adding components as long as the reduced $\chi^2$ increases by more than 10\%, as in \citet{Speranza24}. By doing so we find that the H$_{\beta}$ and [OIII] profiles need four Gaussian components of different FWHMs to be reproduced, ranging from {200 to 2300 km~s$^{-1}$ in the case of H$_{\beta}$ and from 250 to 2300 km~s$^{-1}$ for [OIII] (see top panels of Fig. \ref{figA2} and Table \ref{tab:A1}).} We then use the H$_{\beta}$ kinematics as input for the H$_{\alpha}$ fit, and those of [OIII] for [NII], to help reducing degeneracy. By doing so, we find a blue excess in the H$_{\alpha}$+[NII] profile that cannot be accounted for by the four components (see bottom panels of Fig. \ref{figA2}). In fact, the fit improves by {86\%} if we include a fifth component for H$_{\alpha}$, of {FWHM=3400$\pm$100 km~s$^{-1}$ and blueshifted by -1000$\pm$70} km~s$^{-1}$ (black dotted line in the bottom right panel of Fig. \ref{figA2}). We repeated the same procedure with the original SDSS spectrum, without subtracting the stellar continuum {(i.e. the spectrum shown in Fig. \ref{figA1})}, and {we find the same number of Gaussians and similar kinematics: FWHM=3200$\pm$100 km~s$^{-1}$ and v$_s$=-1400$\pm$100 km~s$^{-1}$ for the fifth component of H$_{\alpha}$.}

\begin{table}
\small
  \caption{Main properties of the H$_{\beta}$, [OIII], and H$_{\alpha}$+[NII] fits from the continuum-subtracted SDSS spectra of J1010, shown in the bottom right panel of Fig. \ref{figA2}.} 
  \centering
\begin{tabular}{l c c c}        
\hline  
  Line   & FWHM & $\rm v_{s}$ & Line flux $\times$ $\rm 10^{15}$ \\ 
 & (km~s$^{-1}$) & (km~s$^{-1}$) & ($\rm erg~cm^{-2}~s^{-1}$) \\ 
\hline
H$_{\beta}$ & &  &  \\ 
 (n)      & 300$\pm$60  &  50$\pm$50 &  1.6$\pm$0.4 \\ 
 (n)      & 200$\pm$100  &-160$\pm$40 &  0.5$\pm$0.4 \\ 
 (i)      & 800$\pm$100 & -50$\pm$20 &  5.0$\pm$1.0 \\ 
 (b)      &2300$\pm$150 &-380$\pm$80 &  5.0$\pm$1.0 \\ 
 \hline
  ${\rm [OIII]}$  & &  &  \\ 
 (n)      & 350$\pm$20  &  70$\pm$10  & 13$\pm$1 \\ 
 (n)      & 250$\pm$20  & -180$\pm$20 & 3$\pm$1 \\ 
 (i)      & 1000$\pm$40 & -10$\pm$5   & 39$\pm$2 \\ 
 (b)      & 2300$\pm$150& -15$\pm$20  & 24$\pm$4 \\ 
\hline
  ${\rm [NII]}$  & &  &  \\ 
 (n)      & 350  &  70  & 7.1$\pm$0.1 \\ 
 (n)      & 250  & -180 & 1.5$\pm$0.2 \\ 
 (i)      & 1000 & -10  & 8.2$\pm$0.5 \\ 
 (b)      & 2300 & -15  & 5.4$\pm$0.4 \\ 
 \hline  
H$_{\alpha}$ & &  &  \\ 
 (n)      & 300 &  50 &   7.1$\pm$0.1 \\ 
 (n)      & 200 &-160 &   2.0$\pm$0.1 \\ 
 (i)      & 800 & -50 &    18$\pm$1   \\ 
 (b)      &2300 &-380 &    22$\pm$2   \\ 
 (b)      &3400$\pm$100 & -1000$\pm$70 & 24$\pm$2 \\
\hline 
 \end{tabular}
 \label{tab:A1}
\tablefoot{Measurements without errors correspond to parameters that have been fixed. The columns report the FWHM, the velocity shift ($\rm v_s$), and the integrated flux obtained from the fits with several Gaussian components 
including narrow (n), intermediate (i), and broad (b). Velocity shifts are relative to the systemic velocity, calculated from the flux weighted wavelength between the two narrow [OIII] peaks, as in \citet{Speranza24}.}
\end{table}

\FloatBarrier
\section{Rotational diagrams}\label{appendixc}

\begin{figure}
   \centering
  {\par\includegraphics[width=0.45\columnwidth]{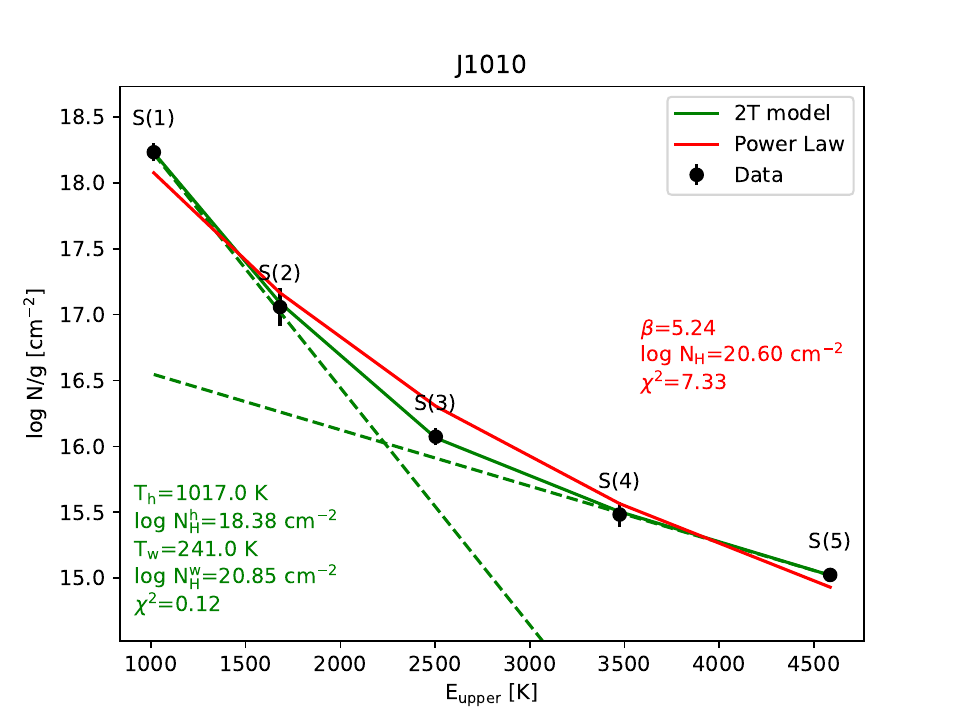}
  \includegraphics[width=0.45\columnwidth]{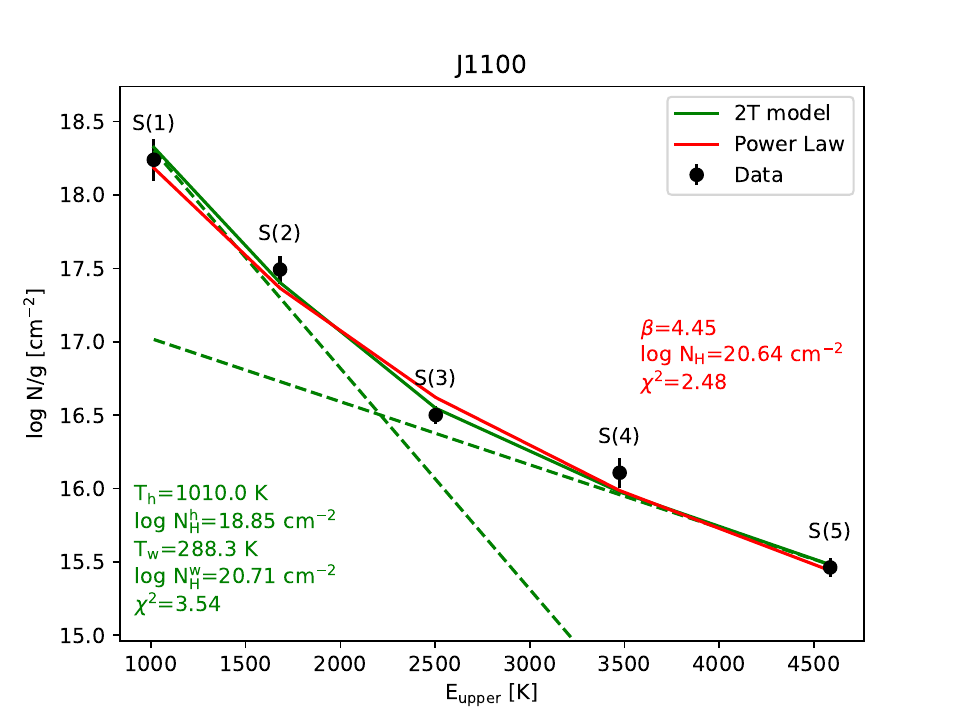}
  \includegraphics[width=0.45\columnwidth]{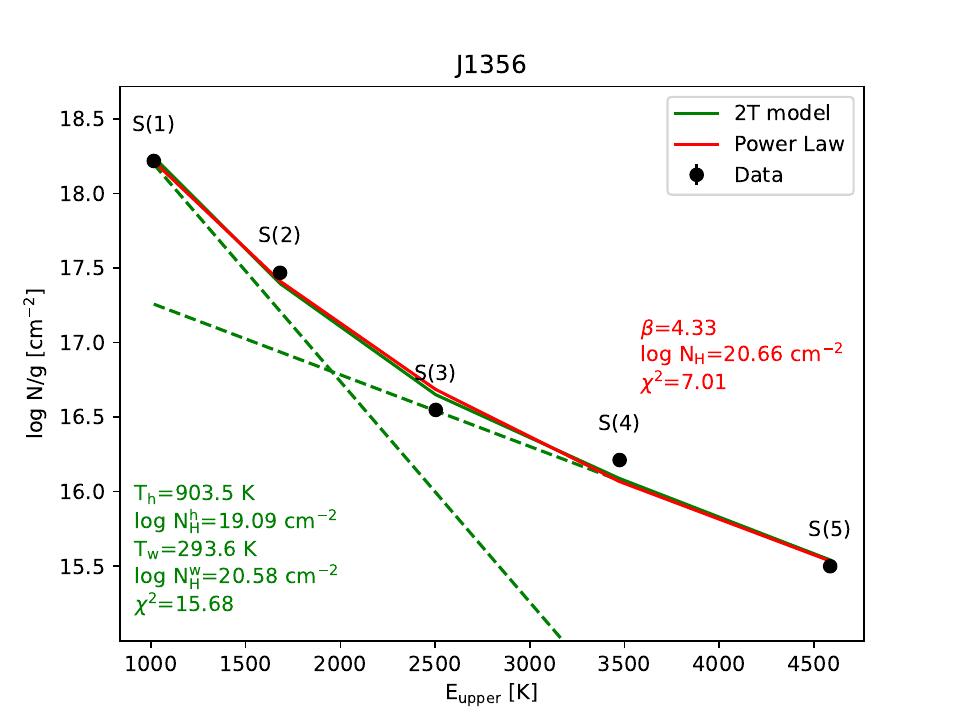}
  \includegraphics[width=0.45\columnwidth]{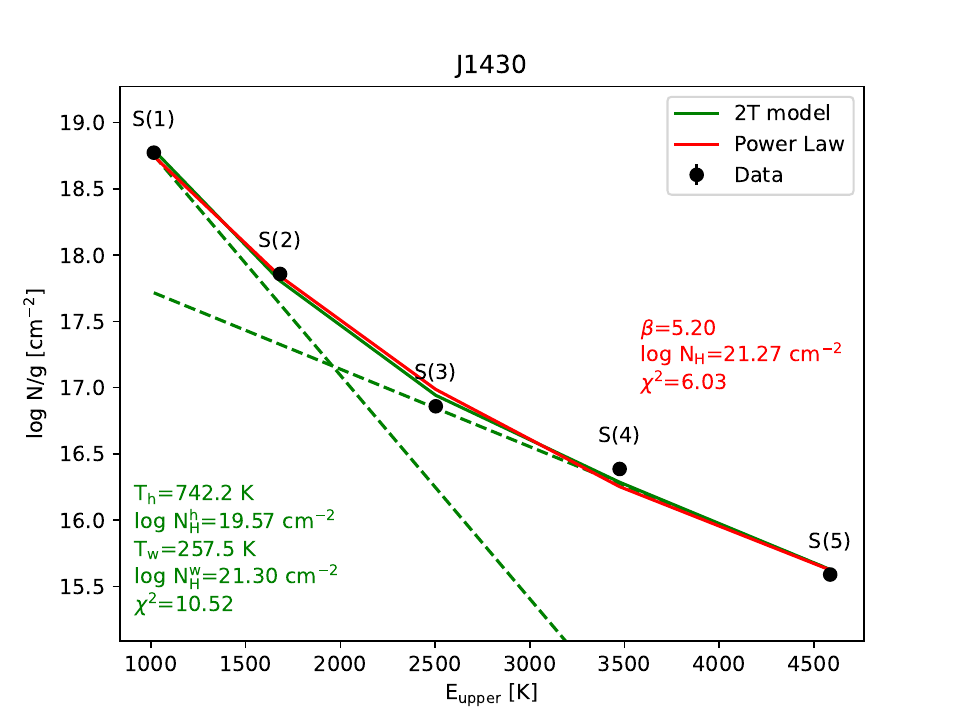}
  \includegraphics[width=0.45\columnwidth]{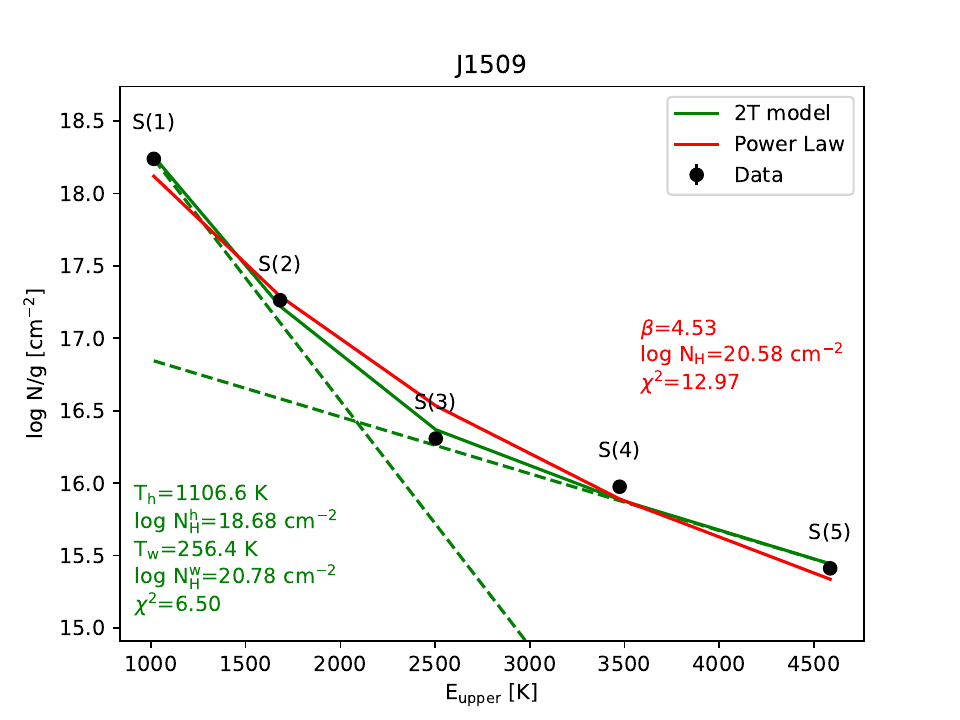}\par}
   \caption{Rotational diagrams of the nuclear region of the QSO2s. The red and green solid lines correspond to the fits with a single power-law (PL) and a two-temperature (2T) model, respectively. The green dashed lines correspond to the warm (T<500 K) and hot (T>500 K) components of the 2T model.}
              \label{figB1}
    \end{figure}

In Fig. \ref{figB1} we show the rotational diagrams of the five QSO2s using the H$_2$ lines from 0-0S(5) to 0-0S(1), whose luminosities and kinematics are reported in Table \ref{tab:lines1}.

\FloatBarrier

\begin{table*}
\small
     \caption{Luminosities and kinematics of the H$_2$ lines.}
         \label{tab:lines1}
\centering                          
\begin{tabular}{lcccccc}        
\hline  
\hline
  Line      & L$_{\rm H_2}$  & FWHM   & $\rm v_{s}$  & L$_{\rm H_2}$     & FWHM    & $\rm v_{s}$ \\  
       &  (10$^{40}$ erg~s$^{-1}$) & (km~s$^{-1}$) & (km~s$^{-1}$) &  (10$^{40}$ erg~s$^{-1}$) & (km~s$^{-1}$) & (km~s$^{-1}$) \\ 
       \hline
    & \multicolumn{3}{c}{J1010} & \multicolumn{3}{c}{J1100}  \\ 
\hline
S5 & {1.82$\pm$0.12} &  341$\pm$26	 &     -26  $\pm$    11 &  {5.30$\pm$0.55}   & 511$\pm$58   &  -7   $\pm$      25   \\
S4 & {0.59$\pm$0.12} &  328$\pm$69	 &      53  $\pm$    32 &  {2.62$\pm$0.56}   & 512$\pm$102  &   32  $\pm$      60   \\
S3 & {1.79$\pm$0.20} &  288$\pm$41	 &      29  $\pm$    17 &  {5.08$\pm$0.48}   & 533$\pm$57   &   72  $\pm$      23   \\
S2 & {1.04$\pm$0.34} &  277$\pm$116  &     -105             &  {3.00$\pm$0.58}   & 327$\pm$66   &  -54  $\pm$      33   \\
S1 & {4.52$\pm$0.56} &  505$\pm$66	 &     -6   $\pm$    27 &  {4.87$\pm$1.53}   & 351$\pm$135  &  -38  $\pm$      58   \\
\hline
    & \multicolumn{3}{c}{J1356} & \multicolumn{3}{c}{J1430}  \\ 
    \hline
S5  & 8.98$\pm$0.17 & 466  $\pm$    9	    & -10  $\pm$  4   &{5.00$\pm$0.09} &  500$\pm$10  &     20$\pm$4     \\
S4  & 5.17$\pm$0.23 & 565  $\pm$    30	    &  2   $\pm$  12  &{3.50$\pm$0.12} &  594$\pm$20  &     31$\pm$8     \\
S3  & 8.79$\pm$0.10 & 507  $\pm$    6	    & -8   $\pm$   3  &{8.16$\pm$0.09} &  557$\pm$6	 &      0$\pm$3  	\\
S2  & 4.39$\pm$0.20 & 573  $\pm$    31      & -28  $\pm$  12  &{4.89$\pm$0.21} &  536$\pm$26  &     58$\pm$11    \\
S1  & 7.17$\pm$0.31 & 604  $\pm$    31      & -11  $\pm$  12  &{11.7$\pm$0.44} &  484$\pm$20  &     16$\pm$9     \\
\hline
    & \multicolumn{3}{c}{J1509} & \multicolumn{3}{c}{}  \\ 
    \hline
S5  &   {5.91$\pm$0.16}     &  394$\pm$12  &      25  $\pm$	   5   & & &\\
S4  &   {2.42$\pm$0.13}     &  415$\pm$30  &      32  $\pm$	   10  & & &\\
S3  &   {4.08$\pm$0.08}     &  361$\pm$8   &      3  $\pm$  3	   & & &\\
S2  &   {2.21$\pm$0.11}     &  309$\pm$17  &     -5	$\pm$	   8   & & &\\
S1  &   {6.07$\pm$0.22}     &  313$\pm$14  &      8	$\pm$	   6   & & &\\
\hline 
\end{tabular}
\tablefoot{Same as in Table \ref{tab:kin}, but for the H$_2$ 0-0S(1), S(2), S(3), S(4), and S(5) emission lines detected in the nuclear MIRI/MRS spectra of the QSO2s, corrected from extinction.}
\end{table*}

\end{appendix}

\end{document}